\documentclass[a4paper]{cas-dc}
\usepackage[authoryear,round]{natbib}
\usepackage{amsmath, amssymb}
\usepackage{bm}
\usepackage{cases}
\usepackage{algorithm}
\usepackage{threeparttable} % 加入导言区
\usepackage{algpseudocode}
\usepackage{multirow}
\usepackage{graphicx}
\usepackage{float}
\usepackage{subcaption}
\usepackage{multirow}
\usepackage{booktabs}
\usepackage{pdflscape}
\renewcommand{\vec}[1]{\bm{#1}} 
\def\tsc#1{\csdef{#1}{\textsc{\lowercase{#1}}\xspace}}
\tsc{WGM}
\tsc{QE}

\begin{document}
\begin{sloppypar}
\let\WriteBookmarks\relax
\def\floatpagepagefraction{1}
\def\textpagefraction{.001}
\shorttitle{Collaborative Learning for Unsupervised Multimodal Image Registration}
\shortauthors{Xiaochen Wei et~al.}

\title{Collaborative Learning for Unsupervised Multimodal Remote Sensing Image Registration: Integrating Self-Supervision and MIM-Guided Diffusion-Based Image Translation}

\author[1,2,3]{Xiaochen Wei}%[<options>]

\author[4]{Weiwei Guo}%[<options>]
\cormark[1]
\ead{weiweiguo@tongji.edu.cn}

\author[1,2,3]{Wenxian Yu}%[<options>]

% Address/affiliation
\affiliation[1]{organization={Shanghai Key Laboratory of Intelligent Sensing and Recognition},
            city={Shanghai},
            postcode={200240}, 
            country={China}}
\affiliation[2]{organization={School of Sensing Science and Engineering},
            city={Shanghai},
            postcode={200240}, 
            country={China}}
\affiliation[3]{organization={Shanghai Jiao Tong University},
            city={Shanghai},
            postcode={200240}, 
            country={China}}
\affiliation[4]{organization={Center for Digital Innovation, Tongji University},
            city={Shanghai},
            postcode={200092}, 
            country={China}}

\cortext[1]{Corresponding author}

\begin{abstract}
The substantial modality-induced variations in radiometric, texture, and structural characteristics pose significant challenges for the accurate registration of multimodal images. While supervised deep learning methods have demonstrated strong performance, they often rely on large-scale annotated datasets, limiting their practical application. Traditional unsupervised methods usually optimize registration by minimizing differences in feature representations, yet often fail to robustly capture geometric discrepancies, particularly under substantial spatial and radiometric variations, thus hindering convergence stability. To address these challenges, we propose a \textbf{Co}llaborative \textbf{L}earning framework for Unsupervised Multimodal Image \textbf{Reg}istration, named \textbf{\textit{CoLReg}}, which reformulates unsupervised registration learning into a collaborative training paradigm comprising three components: (1) a cross-modal image translation network, MIMGCD, which employs a learnable Maximum Index Map (MIM) guided conditional diffusion model to synthesize modality-consistent image pairs; (2) a self-supervised intermediate registration network which learns to estimate geometric transformations using accurate displacement labels derived from MIMGCD outputs; (3) a distilled cross-modal registration network trained with pseudo-label predicted by the intermediate network. The three networks are jointly optimized through an alternating training strategy wherein each network enhances the performance of the others. This mutual collaboration progressively reduces modality discrepancies, enhances the quality of pseudo-labels, and improves registration accuracy. Extensive experimental results on multiple datasets demonstrate that our ColReg achieves competitive or superior performance compared to state-of-the-art unsupervised approaches (e.g., Alto, SCPNet, SSHNet, and SSHNet-D) and even surpasses several supervised baselines (e.g., DHN, MHN, and ReDFeat).
\end{abstract}

\begin{highlights}
\item The accurate registration of multimodal images remains challenging due to substantial variations in radiometry, texture, and structure, which degrade geometric consistency. While supervised deep learning methods have demonstrated strong performance, they often rely on large-scale annotated datasets. Existing unsupervised methods typically rely on minimizing feature differences but struggle to capture accurate geometric transformations under large modality gaps and lack convergence stability.

\item Instead of the conventional "image translation and mono-modal image registration" pipeline, we propose CoLReg, a collaborative learning framework that reformulates unsupervised registration as a mutual reinforcement learning process among three components: an image-to-image translation based on a conditional diffusion model, a self-supervised intermediate registration network, and a distilled cross-modality registration network.

\item We propose a novel MIM-guided conditional diffusion model (MIMGCD) to generate structure-preserving target-domain images for improved registration, addressing the limitations of conventional translation-based pipelines.

\item Extensive experiments on multiple datasets demonstrate that CoLReg not only outperforms state-of-the-art unsupervised methods but also surpasses several supervised baselines.

\end{highlights}

%\nocitep{*}

% Keywords
% Each keyword is separated by \sep
\date{}

\maketitle
\definecolor{method3}{HTML}{F2F2F2}
\definecolor{method1}{HTML}{F7F4ED}  % 淡米色
\definecolor{method0}{HTML}{E6F0EE}  % 浅青灰
\definecolor{method2}{HTML}{E3ECF6}  % 浅紫灰
\section{Introduction}

Multimodal remote sensing image registration establishes the geometric correspondence between images from different sensors or imaging modalities through estimating a global or local geometric transformation, facilitating effective matching and fusion of features across modalities and enhancing the accuracy and reliability of subsequent tasks. Whether for change detection ~\cite{Lv2023NovelEU}, multimodal fusion~\cite{Zhou2024AGS}, object detection~\cite{Xiao2024DGFNetDC}, or visual geo-localization~\cite{Xiao2024STHNDH}, registration accuracy directly affects the quality of these results, making it a fundamental component of multimodal image processing.

The inherent differences in imaging mechanisms and viewing geometries lead to substantial variations in radiometry, texture, gradient, and geometry among multimodal images, thereby posing significant challenges for accurate multimodal image registration. Recent deep learning registration approaches have demonstrated their superiority in handling large geometric differences and modality gaps, compared to traditional ones based on handcrafted features~\cite {Deng2022ReDFeatRD, zhu2024mcnet, Li2019RIFTMI, Dang2015SRIFSA}.

However, deep supervised registration methods typically require large-scale datasets comprising image pairs with precise geometric correspondences between the source and target images. In practice, such ground-truth displacements are often unavailable and must be manually obtained by annotating a substantial number of control points, rendering the dataset preparation process both time-consuming and labor-intensive.

\begin{figure*}[!htb]
    \centering
    \includegraphics[width=\textwidth]{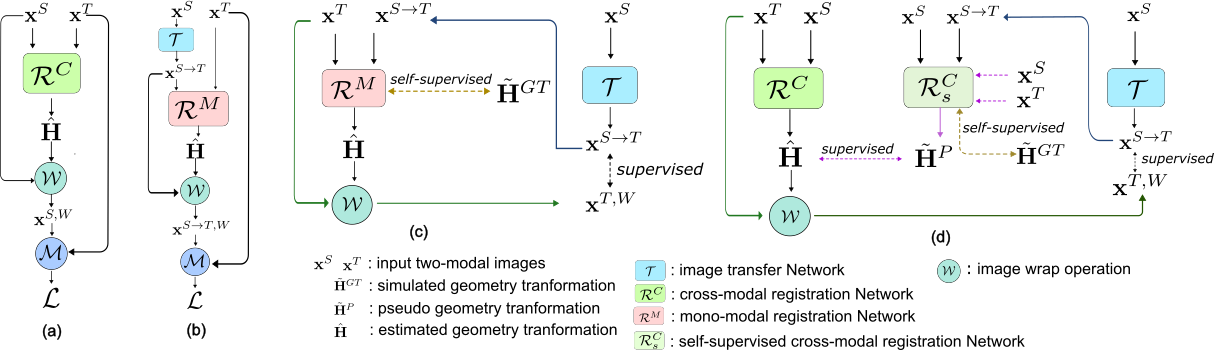}
    \caption{Unsupervised multimodal image registration methods in early research. (a) registration methods based on cross-modality measure; (b) registration methods based on image-to-image translation; (c) split registration learning with image-to-image translation and mono-modal registration; (d) our direct alternative registration learning method based on image-to-image translation and cross-modality registration learning.}
    \label{in}
\end{figure*}

To overcome the limitations of deep supervised methods, various unsupervised approaches have been proposed. As shown in Fig.~\ref{in}, current unsupervised multimodal image registration methods can be broadly classified into three main categories. The first category of methods~\cite{9758703,kong2023indescribable,song2024unsupervised, Koguciuk2021PerceptualLF}, as shown in Fig. \ref{in}(a),  directly learns a registration network $\mathcal{R}^{C}$ with input two modal images $\vec{x}^{S}$ and $\vec{x}^{T}$ in an unsupervised manner by optimizing a dedicated image similarity metric $\mathcal{M}$ between the warped source image and target image. Such approaches rely on designing robust cross-modal similarity metrics that can guide the network in the absence of ground-truth geometric correspondences. For example, Alto\cite{song2024unsupervised} projected the cross-modality images into a modality-invariant feature space for the similarity metric $\mathcal{M}$ calculation. To address the modality gap, the second category~\cite{arar2020unsupervised, chen2022unsupervised, Xu2022RFNetUN}, as shown in Fig.~\ref{in}(b), employs an image-to-image translation network $\mathcal{T}$ that converts one modality into the other. A mono-modality registration network $\mathcal{R}^{M}$ is then utilized, guided by a corresponding similarity metric. In these methods, both the registration and translation networks are trained in an indirect manner, guided by the similarity metric rather than explicit geometric supervision. Registration accuracy relies heavily on the content consistency between images or their corresponding feature representations. When there are significant modality variations, the similarity metrics computed either in the image or feature space often fail to robustly capture the underlying geometric discrepancies, thereby limiting the registration performance. To address this issue, as shown in Fig.~\ref{in}(c), Yu et al.~\cite{yu2024internet} proposed a direct learning method that reformats the cross-modal registration learning into two subproblems comprised of a mono-modal registration network $\mathcal{R}^{M}$ and a translation network $\mathcal{T}$ which are effectively trained via their proposed split optimization strategy. The registration network $\mathcal{R}^{M}$ is directly supervised through synthetically deformed mono-modal image pairs and trained with a direct geometric deformation loss between the simulated geometric transformation $\tilde{\vec{H}}^{GT}$ and the estimation. Furthermore, a lightweight cross-modal registration network is subsequently distilled, with supervision provided by the pseudo-transformations generated by the mono-modal registration network $\mathcal{R}^{M}$ and the translation network $\mathcal{T}$.

However, the "translation $\rightarrow$ mono-registration" framework suffers from several inherent limitations. First, the performance of the mono-modal registration network remains highly sensitive to the extent of the modality gap, which the image translation network $\mathcal{T}$ cannot fully eliminate. For example, simulated mono-modal image pairs such as $\{\vec{x}^{T}, \mathcal{W}(\vec{x}^{T})\}$ and $\{\vec{x}^{S\rightarrow T}, \mathcal{W}(\vec{x}^{S\rightarrow T})\}$ contain no modality gap, while translated pairs $\{\vec{x}^{S\rightarrow T}, \vec{x}^{T}\}$ inevitably retain residual differences, especially under large modality shifts, thereby degrading registration performance. Second, the image translation process may inadvertently discard modality-invariant geometric cues that are critical for accurate registration. Third, the framework comprising both translation and registration networks incurs substantial computational overhead. Although a lightweight cross-modal registration network can be obtained through knowledge distillation, this post hoc distillation inevitably results in performance degradation. While image translation reduces the complexity of the registration task, we advocate that increasing the difficulty of registration learning tasks can instead drive the network to learn more discriminative and modality-invariant features, which are vital for enhancing the robustness and generalizability of registration across diverse cross-modal scenarios.

To address the aforementioned challenges, we propose an unsupervised direct cross-modal image registration learning framework that learns a powerful cross-modal image registration network $\mathcal{R}^{C}$, which we reformat into a collaborative training paradigm involving three key components: an image-to-image translation network $\mathcal{T}$, a self-supervised intermediate cross-modal registration network $\mathcal{R}^{C}_{s}$, and the distilled cross-modal registration network $\mathcal{R}^{C}$, as shown in Fig.~\ref{in} (d). Due to the lack of ground-truth geometric correspondence for direct geometric supervised learning, we employ an intermediate self-supervised cross-modal registration network $\mathcal{R}^{C}_{s}$ which is trained on the synthetic cross-modal image pairs $\{\mathcal{W}(\vec{x}^{S}),\vec{x}^{S \rightarrow T}\}$ generated by the translation network $\mathcal{T}$, along with the simulated the geometric transformation $\tilde{H}^{GT}$. The cross-modal registration network $\mathcal{R}^{C}$ is then trained on the cross-modal image pairs $\{\vec{x}^{S},\vec{x}^{T}\}$ , using the pseudo-labels produced by the intermediate network $\mathcal{R}^{C}_{s}$ as supervision. Among the collaborative training process, the warped image $\vec{x}^{T, W}$, obtained via the estimated geometric transformation $\hat{\vec{H}}$, serves as a supervisory signal to guide the training of the translation network $\mathcal{T}$. To better generate the cross-modal image pairs that stably capture the modal variations, we introduce a \textbf{M}aximum \textbf{I}ndex \textbf{M}ap (MIM)-\textbf{G}uided \textbf{C}onditional \textbf{D}iffusion model (MIMGCD) for unsupervised cross-modal image translation. Maximum Index Map (MIM), which is a classic modal-invariant feature, is employed to guide the reverse diffusion process, thereby ensuring that the translated images retain richer structural and geometric details essential for accurate registration.

Compared to image translation and mono-modal registration learning, our cross-modal registration framework explicitly encourages the network to focus on modality-invariant geometric information for accurate registration during training. This design not only mitigates the potential loss of geometry-relevant features often introduced by the image translation network but also enables the model to adapt to varying levels of modality discrepancy. Furthermore, the collaborative training strategy allows the three components to reinforce each other, ultimately enabling the learning of a powerful cross-modal registration network without requiring explicit modality translation at inference.

The main contributions of this paper are summarized as follows:
\begin{itemize}
\item We propose a novel unsupervised cross-modal image registration framework, reformulated as a collaborative learning paradigm that integrates three key components: a cross-modal image translation network, an intermediate self-supervised registration network, and a distilled cross-modal registration network. Extensive experiments demonstrate the superiority of the proposed framework over existing state-of-the-art methods.

\item We design an alternating optimization strategy to ensure stable and effective training of the three components. Specifically, the intermediate cross-modal registration network $\mathcal{R}^{C}_{s}$ is trained on the synthetic cross-modal image pairs generated by the image translation network $\mathcal{T}$ along with the simulated geometric transformations; the cross-modal registration network $\mathcal{R}^{C}$ is subsequently trained using the pseudo-labels produced by the intermediate network $\mathcal{R}^{C}_{s}$, while the image translation network $\mathcal{T}$ is further optimized using the cross-modal image pairs along with the geometric transformation estimated by the registration network $\mathcal{R}^{C}$. The three networks are trained in an alternating manner and mutually reinforce each other throughout the whole training.

\item We introduce a novel MIM-guided conditional diffusion model (MIMGCD) for unsupervised image-to-image translation. Leveraging the modality-invariant characteristics of the Maximum Index Map (MIM), the reverse diffusion process is guided to preserve structural and geometric cues essential for registration. This guidance enables the generation of high-fidelity target-domain images with detailed information beneficial for downstream cross-modal registration.

\end{itemize}

\section{Related Works}
\subsection{Supervised Deep Learning for Image Registration}
For multimodal image registration, early deep learning methods~\cite{Quan2022DeepFC, Cui2022CrossModalityIM} focused on learning modality-invariant discriminative representations of multimodal image patches. These approaches typically enhanced descriptors by modifying network architectures and contrastive loss functions. For example, Cnet~\cite{Quan2022DeepFC} employs an attention-based feature learning network and a novel feature correlation loss function, which emphasizes common features across different modalities. As the need for modality and viewpoint repeatable keypoints arose, researchers began focusing on using deep networks to learn keypoint detectors and descriptors simultaneously.  GRiD~\cite{Liu2024GRiDGR} tackles feature point offset issues by establishing pixel-level correspondences between images, compensating for nonlinear radiometric and geometric distortions in multimodal images. This is achieved through guided refinement, which enhances reference point reliability using a Maximum Index Map and fine localization to optimize matching accuracy. While the methods described above separate feature detection and descriptor learning from transformation parameter optimization, this can lead to error accumulation. To address this, recent approaches have proposed end-to-end networks that directly predict transformation parameters. DHN~\cite{detone2016deep} was one of the first to frame homography parameter estimation as a regression problem, focusing on the displacements of four corner points. DLKFM~\cite{zhao2021deep} builds on this by extending the traditional Lucas-Kanade algorithm, constructing a deep Lucas-Kanade feature map to align multimodal image pairs pixel by pixel. IHN~\cite{cao2022iterative} further improves registration accuracy by designing a trainable iterator that iteratively refines the estimated parameters. To enhance registration performance, RHWF~\cite{cao2023recurrent} integrates homography-guided image warping and the FocusFormer attention mechanism within a recurrent framework, progressively enhancing feature consistency and refining correspondences in a global-to-local fashion. MCNet ~\cite{zhu2024mcnet} reduces computational overhead during iteration by combining a multi-scale strategy with correlation search, in contrast to earlier methods that relied on single-scale iterative refinement.  Lastly, to mitigate the impact of initial estimated parameters, SDME~\cite{pmlr-v235-zhang24ar} uses a multi-task network to predict both sparse and dense features simultaneously, leveraging the strengths of both sparse feature matching and dense direct alignment for improved multimodal image registration across diverse scenes.

\subsection{Unsupervised Deep Learning for Image Registration}
To address the reliance of deep learning methods on accurately labeled deformation ground truth, unsupervised methods have garnered increasing attention from researchers in recent years. Current unsupervised learning methods can be categorized into two types: one focuses on designing unsupervised loss functions, while the other aims to map multimodal images to a common domain. 

MU-Net \cite{9758703} designs a new loss function paradigm based on structural similarity, which uses the structural descriptor of oriented gradient to calculate the similarity between multimodal images. UDHN~\cite{zhang2020content} introduced a novel triplet loss function based on deep features tailored for unsupervised training. Similarly, biHomE ~\cite{Koguciuk2021PerceptualLF} proposed a perceptual loss called bi-HomE to measure the distance between a warped source image and a target image in the feature space. However, due to large nonlinear radiometric differences, these loss functions struggle to capture geometric differences between cross-modal image pairs robustly, hindering network training toward minimizing deformation differences. Alto~\cite{song2024unsupervised} adopts Barlow Twins Loss to address modality gaps and propose an extended version, Geometry Barlow Twins Loss, to handle geometry gaps. 

Unlike methods based on cross-modal loss functions, image translation or mapping-based methods reduce the radiometric differences between multimodal images by translating the source image into the domain of the target image or mapping both images into a common domain. DFMIR~\cite{chen2022unsupervised} translates the multi-modal images into monomodal by using a discriminator-free translation network without adversarial loss and a patchwise contrastive loss. INNReg \cite{guo2025unsupervised} proposes an Invertible Neural Network (INN) with a dynamic convolution-based local attention mechanism and a novel barrier loss function to translate multi-modal images into mono-modal ones for accurate alignment. SCPNet~\cite{zhang2025scpnet} applies intra-modal self-supervised learning to map cross-modality image pairs into a consistent feature domain. SHHNet~\cite{yu2024internet} integrates interleaved modality transfer and self-supervised homography prediction, using an alternating optimization strategy to enhance both components while employing a fine-grained homography feature loss and distillation technique for improved accuracy and generalization.
\subsection{Unsupervised Multimodal Image-to-Image Translation}
Unsupervised multimodal image-to-image translation (UMIT) aims at translating images between different modalities without relying on paired training data. CycleGAN~\cite{CycleGAN2017} is one of the earliest representative methods, which introduced cycle consistency loss to enforce consistency between the generated and original images, allowing unpaired image translation. MUNIT~\cite{Huang2018MultimodalUI} builds on the shared latent space assumption, where images from different modalities are mapped to a common latent representation. It further decouples image content and style, allowing for diverse outputs by sampling different style codes while maintaining content consistency. Recently, diffusion-based UMIT methods have achieved significant improvements in generation quality and stability. UNIT-DDPM~\cite{Sasaki2021UNITDDPMUI} learns the joint distribution using Markov chains and Denoising Markov Chain Monte Carlo (MCMC) methods, achieving high-quality image generation; Unpaired Neural Schrödinger Bridge (UNSB)~\cite{kim2023unsb} utilizes the Schrödinger Bridge framework to model stochastic differential equations (SDEs) for distribution transformation, adapting to high-resolution images; ContourDiff combines structural consistency constraints in medical image translation, preserving boundaries and fine details, further improving the generation quality; FDDM~\cite{lifddm} enhances medical image translation by using a frequency-decoupling strategy, focusing on frequency domain information to optimize the balance between spatial and frequency features of the image.

\section{Methodology}
\subsection{The Overall Framework of Unsupervised Multimodal Image Registration}
Given the image pair $\{(\vec{x}^{S}, \vec{x}^{T})\}$ from two modalities with geometric deformation between them, the supervised cross-modal image registration network $\mathcal{R}^{C}$ can be learned to estimate the geometric transformation with the the loss formulated as:
\begin{equation}
  \label{e1}
    \varPhi_{C} = \arg\min_{\varPhi} \, \mathcal{L}_{c}\left(\mathcal{R}^{C}(\vec{x}^S,\vec{x}^T;\varPhi),\vec{H}^{GT}\right)
\end{equation}

Where $\mathcal{L}_{c}$ is the loss between the estimated transformation matrix and the ground truth $\vec{H}^{GT}$, $\varPhi$ is the registration network parameters to be optimized. However, in practice, the ground truth $\vec{H}^{GT}$ is often unavailable, especially for the cross-modal images. Therefore, in this paper, we aim to learn the registration network in an unsupervised manner. We present a pseudo-label-based multi-model co-training framework, which simultaneously conducts self-supervised training of an intermediate registration network to generate pseudo ground-truth transformation matrices while training a cross-modal registration network using these pseudo-labels as supervision. This multi-model co-training framework enables the pseudo-labeling generation and the registration network training processes to mutually reinforce each other as the training process goes on.

Specifically, we employ an intermediate registration network $\mathcal{R}^{C}_{s}(\varPhi_s)$ that is self-trained on simulated cross-modal image pairs generated by image translation network $\mathcal{T}(\varPhi_t)$, along with simulated geometric transformations $\vec{H}^{GT}$, and the registration network $\mathcal{R}^{c}(\varPhi_c)$ is then trained using the pseudo labels $\vec{H}^{PL}$ produced by the $\mathcal{R}^{C}_{s}$. This process can be formulated as
\begin{equation}
\begin{cases}
    \begin{aligned}
        \hat{\varPhi}_{C} &= \arg\min_{\varPhi_C} \, \mathcal{L}_{c}\left(\mathcal{R}^{C}(\vec{x}^S,\vec{x}^T;\varPhi),\vec{H}^{PL}\right)\\
        \vec{H}^{PL} &= \mathcal{R}^{C}_s(\vec{x}^S,\vec{x}^T;\varPhi_s)
    \end{aligned}
\end{cases}
\end{equation}
where the intermediate registration network $\mathcal{R}^{C}_{s}(\varPhi_s)$ is trained on the  simulated data $\{(\vec{x}^{S'}, \vec{x}^{S\rightarrow T}, \vec{H}^{GT})\}$ with the objective
\begin{equation}
\begin{cases}
        \begin{aligned}
        \mathcal{L}_{s}(\varPhi_s)&=\mathcal{L}_{s}\left(\mathcal{R}^{C}_{s}(\vec{x}^{S'}, \vec{x}^{S\rightarrow T},\varPhi_s),  \vec{H}^{GT}\right),\\
        \vec{x}^{S'} &= \mathcal{W}_{\vec{H}^{GT}}(\vec{x}^{S}),\\
        \vec{x}^{S\rightarrow T} & = \mathcal{T}(\vec{x}^{S}, \varPhi_t)
    \end{aligned}
\end{cases}
\end{equation}
where we need train a image translation network $\mathcal{T}(\varPhi_t)$. Due to the lack of well aligned image pairs $(\vec{x}^S, \vec{x}^T)$ in the unsupervised setting, we utlize the estimated $\hat{H}$ to wrap the $\vec{x}^{T}$, yielding the cross-modal image pairs $\{(\vec{x}^{S}, \vec{x}^{T,W})\}$ to train the image translation network $\mathcal{T}(\varPhi_t)$, formulated as 
\begin{equation}
    \begin{cases}
        \begin{aligned}
            \mathcal{L_t}(\varPhi_t)&=\mathcal{L}_t(\mathcal{T}(\vec{x}^{s}, \varPhi_t), \vec{x}^{T,W}),\\
            \vec{x}^{T,W} & = \mathcal{W}_{\hat{H}}(\vec{x}^{T})
        \end{aligned}
    \end{cases}
\end{equation}
where $\mathcal{W}(\cdot)$ is the warp operation.

\begin{figure}[!htbp]
    \centering
    \includegraphics[width=\linewidth]{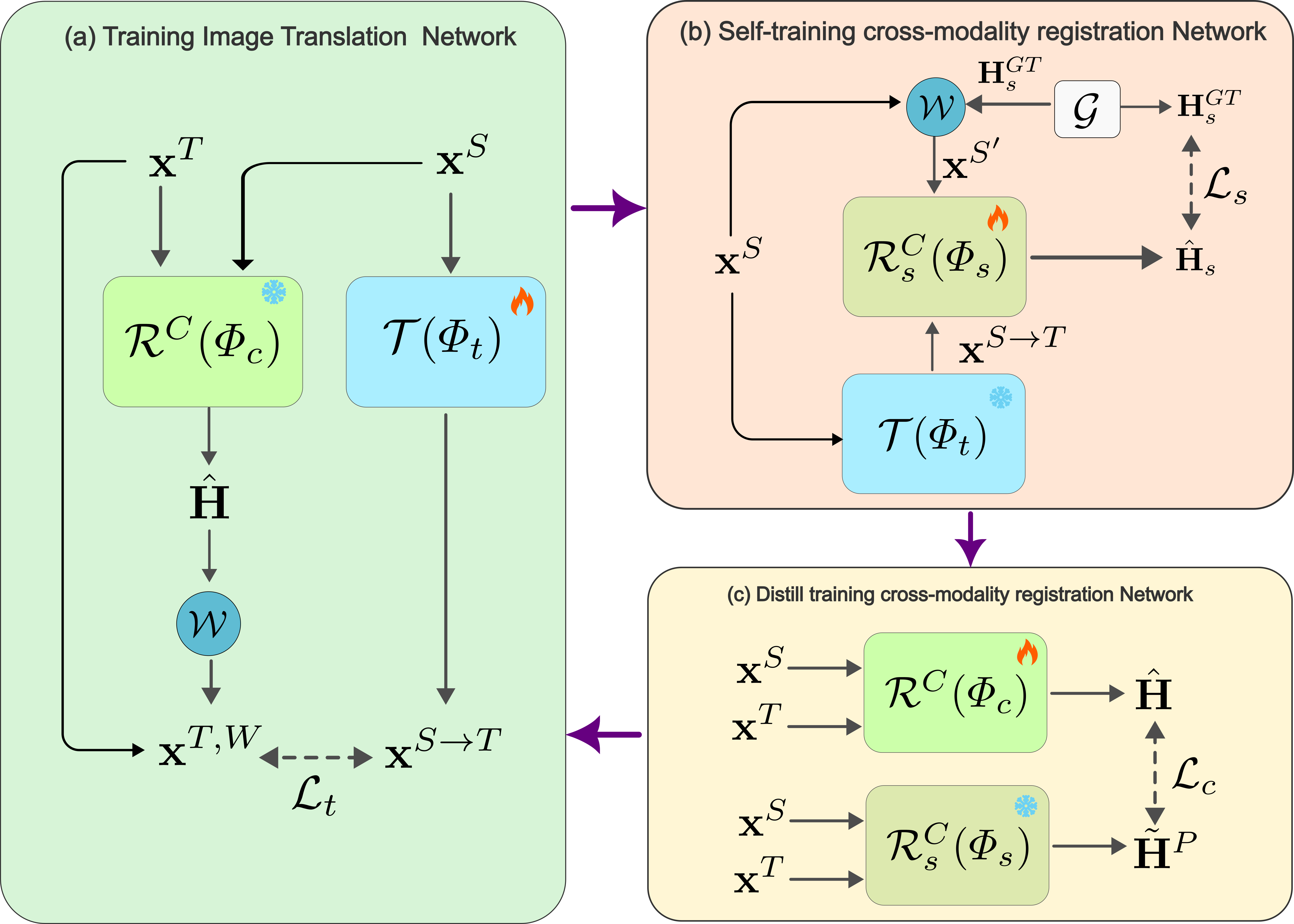}
    \caption{
    Overview of alternating optimization in CoLReg 
    (Collaborative Learning for Unsupervised Multimodal Image Registration). 
    (a) Optimize MIMGCD $\mathcal{T}(\varPhi_t)$; 
    (b) Optimize SRegNet $\mathcal{R}_s^C(\varPhi_s)$; 
    (c) Optimize URegNet $\mathcal{R}^C(\varPhi_c)$.
    }
    \label{overview}
\end{figure}

Since the training involves multi-objective losses $\mathcal{L}_{c}, \mathcal{L}_s, \mathcal{L}_{t}$ optimization, we adopt an alternating optimization scheme to ensure stable convergence and mutual reinforcement. The proposed framework is illustrated in Fig.~\ref{overview}. For the cross-image translation network, we present a MIM-guided conditional diffusion model (MIMGCD) consisting of two MIM encoders, $\Psi_{\mathrm{MIM}}^s(\varPhi_{\mathrm{mim}}^s)$ and $\Psi_{\mathrm{MIM}}^t(\varPhi_{mim}^t)$, along with a conditional diffusion model, $\Psi_{\mathrm{diff}}(\varPhi_{\mathrm{diff}})$. The trained MIMGCD generates cross-modality image pairs, $\{x^{S^{\prime}},x^{S\rightarrow T}\}$ along with simulated the geometric transformation $H^{GT}$, which are used to train the intermediate registration network  $\mathcal{R}^{C}_{s}(\varPhi_s)$. The trained $\mathcal{R}^{C}_{s}(\varPhi_s)$ then generates the pseudo-label $\vec{H}^{PL}$ to supervise the training of the registration network, $\mathcal{R}^{C}(\varPhi_c)$. Furthermore, the  $\vec{x}^{T}$ is wrapped by an estimated transformation by the trained registration network to guide the training of the image-to-image translation network, thereby forming a closed-loop collaborative learning cycle.

\subsection{The Detail of CoLReg}
\subsubsection{MIM-Guided Conditional Diffusion Model for Cross-Modality Image Translation }
\label{mimgcd}
\begin{figure}[!htbp]
    \centering
    \includegraphics[width=85mm]{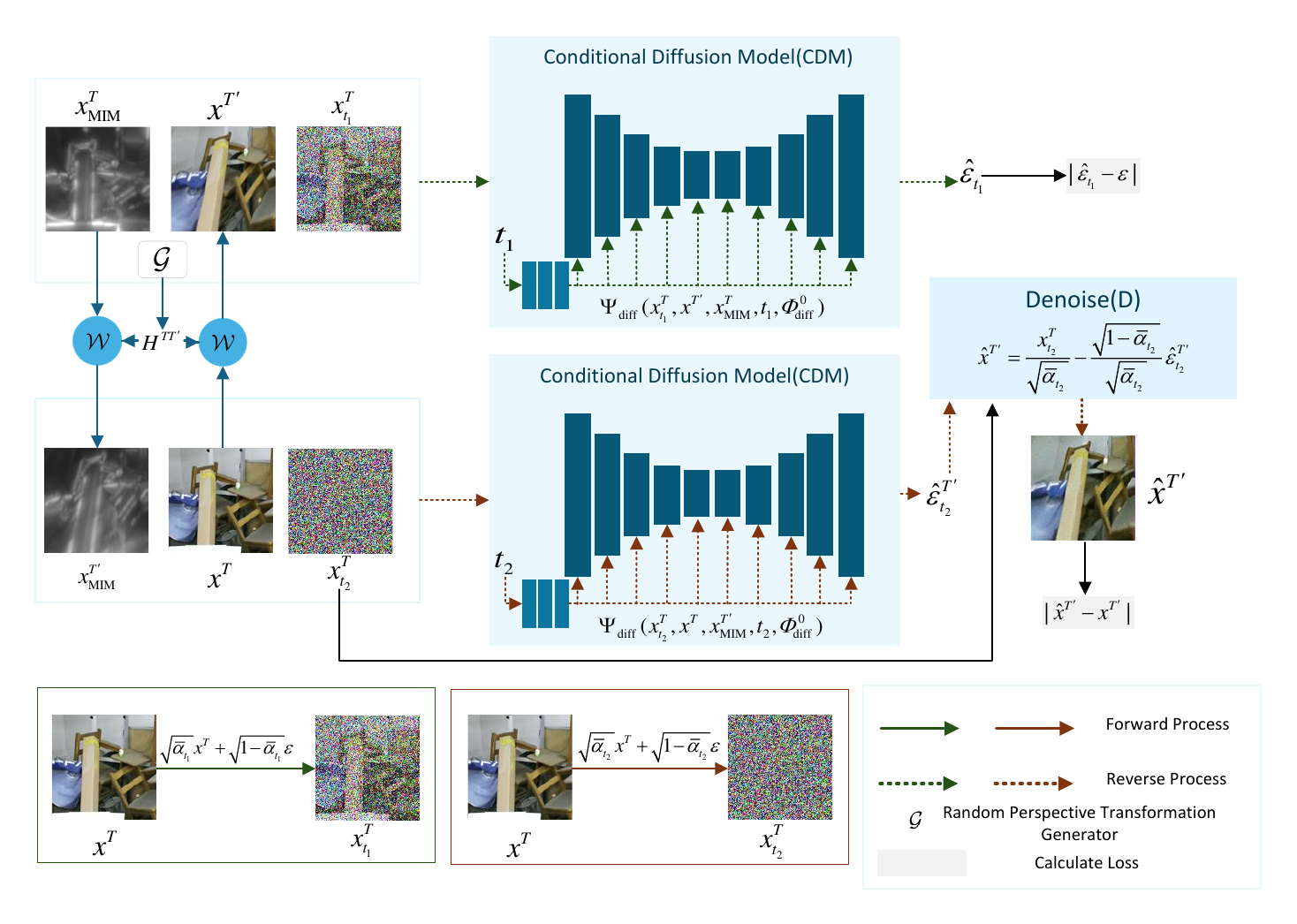}
    \caption{The detail training of $\Psi_{\mathrm{diff}}$ at $it==0$ with two forward processes and reverse processes. The first reverse process 
    $\Psi_{\mathrm{diff}}^0(\vec{x}_{t_1}^T,\vec{x}^{T^{\prime}},\vec{x}_{\mathrm{MIM}}^T)$ is to predict the noise $\hat{\varepsilon}_{t_1}$.
    The second reverse process $\Psi_{\mathrm{diff}}^0(\vec{x}_{t_2}^T,\vec{x}^T,\vec{x}_{\mathrm{MIM}}^{T^{\prime}})$ is to directly obtain the image $\hat{\vec{x}}^{T^{\prime}}$ that is corresponding with condition $\vec{x}_{\mathrm{MIM}}^{T^{\prime}}$.}
    \label{l1}
\end{figure}

\begin{figure*}
    \centering
    \includegraphics[width=170mm]{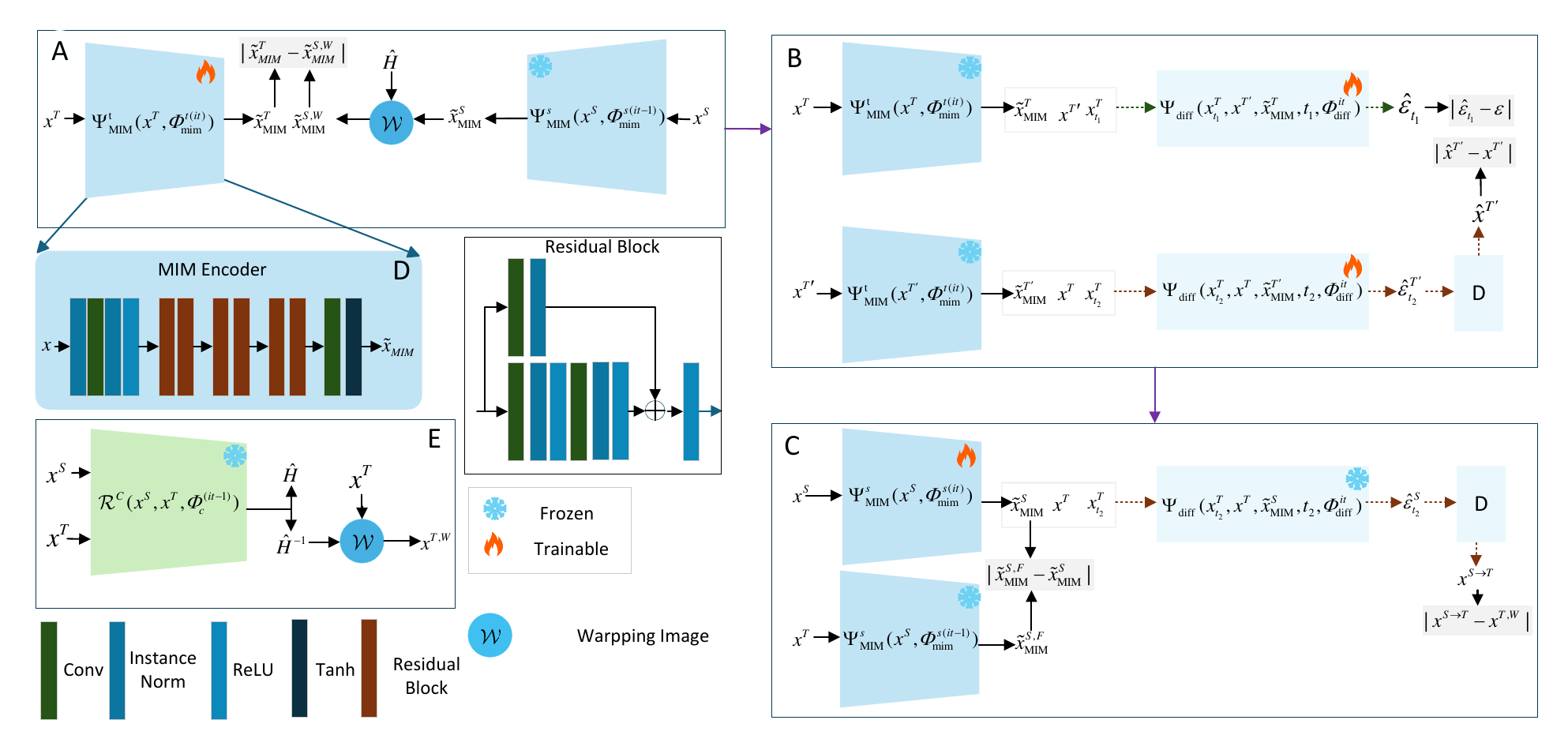}
    \caption{The detail optimization of MIMGCD $\mathcal{T} = \{\Psi_{\mathrm{MIM}}^{t},\Psi_{\mathrm{MIM}}^{s}, \Psi_{\mathrm{diff}}\}(it>=1)$. (A): The training of $\Psi_{\mathrm{MIM}}^{t}(\varPhi_{\mathrm{mim}}^{t(it)})$; (B): The training of $\Psi_{\mathrm{MIM}}^{s}(\varPhi_{\mathrm{mim}}^{s(it)})$; (C): The training of $\Psi_{\mathrm{MIM}}^{s(it)}$; (D): The network architecture of $\Psi_{\mathrm{diff}}(\varPhi_{\mathrm{diff}}^{(it)})$; (E): The obtaining of pseudo-labels $\vec{x}^{T,W}$ and predicted transformations $\hat{H}$ by the $\mathcal{R}^C(\varPhi_{c}^{(it-1)})$.} 
    \label{r2}
\end{figure*}

Existing image-to-image translation models usually require well-aligned cross-modal image pairs for effective training. In the absence of ground-truth supervision, these models often suffer from inaccurate image generation and the loss of critical content and geometric details that are essential for registration. To address these issues, we propose a novel Maximum Index Map(MIM)-Guided Conditional Diffusion Model (MIMGCD), as shown in Fig.~\ref{l1} and~\ref{r2}, which leverages the MIM $\vec{x}_{\mathrm{MIM}}$ or the MIM-like features $\tilde{\vec{x}}_{\mathrm{MIM}}$ extracted via a MIM Encoder $\Psi_{\mathrm{MIM}}$, as a conditioning mechanism to guide the reverse process of the diffusion model. Compared to existing image-to-image translation methods, the proposed MIMGCD offers two key advantages: (1) the MIM-like features exhibit strong modality-invariant properties; (2) the MIM-like features emphasize object boundaries and structural contours, encouraging the network to preserve fine-grained geometric details during training. This ensures that the geometric features of the translated images remain consistent with those of the source images.

To initialize the training of the image translation network, a major problem arises from the lack of the estimated geometric transformation required to wrap the target image to generate the supervised signal $\vec{x}^{T, W}$. To address these issues, in the initial alternating optimization, i.e., when the alternating index $it = 0$, as shown in Fig. ~\ref{l1}, we begin by training the conditional diffusion model $\Psi_{\mathrm{diff}}(\varPhi_{\mathrm{diff}}^0)$ using both the target image $\vec{x}^{T}$ and the corresponding MIM features $\vec{x}^{T}_{\mathrm{MIM}}$ extracted by handcrafted traditional method as input. Since the MIM is modal-invariant, this initialization allows the diffusion model to capture basic modality-invariant generative priors, which provide a stable basis for the subsequent MIM-guided cross-modal translation once geometric supervision becomes available.

Firstly, the two forward processes are to obtain noise images $\vec{x}_{t_1}^{T}$ and $\vec{x}_{t_2}^{T}$ respectively, which are given by:
\begin{equation}
\begin{aligned}
\vec{x}_{t_1}^{T} &= \sqrt{\bar{\alpha_{t_1}}\vec{x}^T} + \sqrt{1-\bar{\alpha_{t_1}}}\varepsilon \\
\vec{x}_{t_2}^{T} &= \sqrt{\bar{\alpha_{t_2}}\vec{x}^T} + \sqrt{1-\bar{\alpha_{t_2}}}\varepsilon
\end{aligned}
\end{equation}
where $\vec{x}^T$ is target image, $t_1$ and $t_2$ are timesteps, $\varepsilon \thicksim \mathcal{N}(0,I)$, and $\bar{\alpha}_{t}(t=t_1,t_2)$ is calculated by:
\begin{equation}
\bar{\alpha}_{t} = \prod_{s=1}^{s=t}1-\beta_{t}
\end{equation}
where $\beta_{t}$ is the noise schedule for controlling the amount of noise added to the image.
Then, as shown in Fig.~\ref{l1},  two reverse processes $\Psi_{\mathrm{diff}}(\vec{x}_{t_1}^T, \vec{x}^{T^{\prime}},\vec{x}_{\mathrm{MIM}}^T,t_1,\varPhi_{\mathrm{diff}}^0)$ and $\Psi_{\mathrm{diff}}^{(0)}(\vec{x}_{t_2}^T,\vec{x}^T, \vec{x}_{\mathrm{MIM}}^{T^{\prime}},t_2)$ are trained to predict noise $\hat{\varepsilon}_{t_1}$ from $x_{t_1}^T$ and obtain clean image corresponding with $\vec{x}_{\mathrm{MIM}}^{T^{\prime}}$ from $\vec{x}_{t_2}^T$. $\vec{x}_{\mathrm{MIM}}^T$ is the MIM of $vec{x}^T$ and calculated by:
\begin{equation}
\begin{aligned}
    \vec{x}_{\mathrm{MIM}}^T(x,y) & = \max({A_o(x,y)}_1^{N_o})\\
    A_o(x,y)  & = \sum_{s=1}^{N_s}A_{so}(x,y)\\
    A_{so}(x,y) & = \sqrt{E_{so}(x,y)^2+O_{so}(x,y)^2}\\
    E_{so}(x,y) &= \vec{x}_{\mathrm{MIM}}^T(x,y)*L^{even}(x,y,s,o)\\
    O_{so}(x,y) &= \vec{x}_{\mathrm{MIM}}^T(x,y)*L^{odd}(x,y,s,o)
\end{aligned}
\end{equation}
where $N_o$ and $N_s$ are the number of orientations and scales respectively, $s$ and $s$ denote the scale index and orientation index of wavelets, $L^{even}$ and $L^{odd}$ are even-symmetric and odd-symmetric wavelets respectively, and $\max$ representations finding the maximum value in orientation dimension.
For the output of first reverse process $\Psi_{\mathrm{diff}}(\vec{x}_{t_1}^T,\vec{x}^{T^{\prime}},\vec{x}_{\mathrm{MIM}}^T,t_1,\varPhi_{\mathrm{diff}}^0)$ $\hat{\varepsilon}_{t_1}$, the loss $\mathcal{L}_{n}$ is calculated as:
\begin{equation}
    \mathcal{L}_{n} = \sum |\hat{\varepsilon}_{t_1}-\varepsilon|
\end{equation}

To speed up image-to-image translation of the diffusion model during inference, following our previous work~\cite{wei2025osdm} for supervised image-to-image translation, we added a one-step image translation reverse process $\Psi_{\mathrm{diff}}(\vec{x}_{t_2}^T,\vec{x}^T, \vec{x}_{\mathrm{MIM}}^{T^{\prime}},t_2,\varPhi_{\mathrm{diff}}^0)$. In this inverse process, we generate the translated image from the target image $x^T$ with a lot of noise added. According to Tweedie's formula, $\hat{\vec{x}}^{T^{\prime}}$ is obtained by one step as:
\begin{equation}
\begin{aligned}
    \hat{\vec{x}}^{T^{\prime}}&= \frac{\vec{x}_{t_2}^{T}}{\sqrt{\bar{\alpha}_{t_2}}} - \frac{\sqrt{1-\bar{\alpha}_{t_2}}}{\sqrt{\bar{\alpha}_{t_2}}} \hat{\varepsilon}^{T^{\prime}}_{t_2}\\
    \hat{\varepsilon}^{T^{\prime}}_{t_2} &= \Psi_{\mathrm{diff}}^{(0)}(\vec{x}_{t_2}^T,\vec{x}^T,\vec{x}_{\mathrm{MIM}}^{T^{\prime}},t_2,\varPhi_{\mathrm{diff}}^0)
\end{aligned}
\end{equation}

The loss of this reverse process is given by:
\begin{equation}
    \mathcal{L}_{t} = \sum |\vec{x}^{T^{\prime}}-\hat{\vec{x}}^{T^{\prime}}|\circ \hat{M}^{T^{\prime}}
\end{equation}
where $\hat{M}^{T^{\prime}}$ is used to mask the padding pixels in $\vec{x}^{T^{\prime}}$.
For $it==0$, the loss $L_{I}^{0}$ of training MIMGCD is given by:
\begin{equation}
    \mathcal{L}_{I}^{0} =   \mathcal{L}_{n} + \mathcal{L}_{t}
\end{equation}

To narrow the domain gap between self-supervised cross-modal image pairs and unsupervised image pairs, when $it>=1$, as shown in Fig. \ref{r2}, we utilize the pseudo-labels $\vec{x}^{T, W}$ and estimated transformation $\hat{H}$ to supervise the training of MIMGCD $\mathcal{T}(\varPhi_{t}^{(it)}) = \{\Psi_{\mathrm{MIM}}^{t}(\varPhi_{\mathrm{mim}}^{t(it)}),\Psi_{\mathrm{MIM}}^{s}(\varPhi_{\mathrm{mim}}^{s(it)}), \Psi_{\mathrm{diff}}(\varPhi_{\mathrm{mim}}^{(it)})\}$. \\

Firstly, as depicted in the Fig. \ref{r2}(A), we train MIM encoder for target images $\Psi_{\mathrm{MIM}}^{t}(\varPhi_{\mathrm{mim}}^{t(it)})$ by the supervision of $\tilde{x}_{\mathrm{MIM}}^{S,W}$. The loss function $\mathcal{L}_{I}^{mt}$ for optimizing the parameters $\varPhi_{\mathrm{mim}}^{t(it)}$ is calculated by:
\begin{equation}
\begin{aligned}
    &\mathcal{L}_{I}^{mt} = \sum |\tilde{\vec{x}}_{\mathrm{MIM}}^{S,W}-\tilde{x}_{\mathrm{MIM}}^{T}| \circ \hat{M}^{ST}\\
    &\tilde{\vec{x}}_{\mathrm{MIM}}^{T} = \Psi_{\mathrm{MIM}}^{t}(\vec{x}^T,\varPhi_{mim}^{t(it)})\\
    &\tilde{\vec{x}}_{\mathrm{MIM}}^{S,W} = \mathcal{W}(\hat{H}, \tilde{\vec{x}}_{\mathrm{MIM}}^{S})
    % & \hat{H}^{ST} = \mathcal{F}_{gw}(P_4,P_4+\Delta \hat{P}_{N_i}^{ST})\\
    % & \Delta \hat{P}_{N_i}^{ST} = \Phi_{reg}^{it-1}(x^S,x^T)
\end{aligned}
\end{equation}
where $\hat{M}^{ST}\in\{0,1\}^{B\times 1 \times H \times W}$ is to mask the padding pixels in $\tilde{\vec{x}}_{MIM}^{S,W}$. For $it==1$, $\tilde{\vec{x}}_{\mathrm{MIM}}^{S}$
is the normalization of $\vec{x}_{\mathrm{MIM}}^{S}$, $\vec{x}^{S,N}_{\mathrm{MIM}}$, which is given by:
\begin{equation}
    \vec{x}_{\mathrm{MIM}}^{S,N} = 2\frac{\vec{x}_{\mathrm{MIM}}^{S}-\min(\vec{x}_{\mathrm{MIM}}^{S})}{\max(\vec{x}_{\mathrm{MIM}}^{S})-\min(\vec{x}_{\mathrm{MIM}}^{S})} - 1
\end{equation}
where the minimum and maximum values are computed separately for each sample and channel.
By minimizing $\mathcal{L}_{I}^{mt}$, the difference between the MIM-like of the source image and the MIM-like of the target image decreases, which helps to utilize $\tilde{\vec{x}}^{S}_{MIM}$ to obtain higher quality translated images $\vec{x}^{S\rightarrow T}$.

Then, as shown in Fig. \ref{r2}(B), we train conditional diffusion model $\Psi_{\mathrm{diff}}$ to update the parameters $\varPhi_{\mathrm{diff}}^{(it)}$ using the self-supervised data. For $it>=1$, we utilize the MIM-likes features $\tilde{\vec{x}}_{\mathrm{MIM}}^{T}$ and $\tilde{\vec{x}}_{\mathrm{MIM}}^{T^{\prime}}$ extracted from $\vec{x}^{T}$ and $\vec{x}^{T^{\prime}}$ by $\Psi_{\mathrm{MIM}}^{t}(\varPhi_{\mathrm{mim}}^{t(it)})$. Except for $\tilde{\vec{x}}_{\mathrm{MIM}}^{T}$ and $\tilde{\vec{x}}_{\mathrm{MIM}}^{T^{\prime}}$, the remaining parts are the same as the training of $\Psi_{\mathrm{diff}}(\varPhi_{\mathrm{diff}}^{(0)})$.

\begin{figure}[!htbp]
    \centering
    \includegraphics[width=85mm]{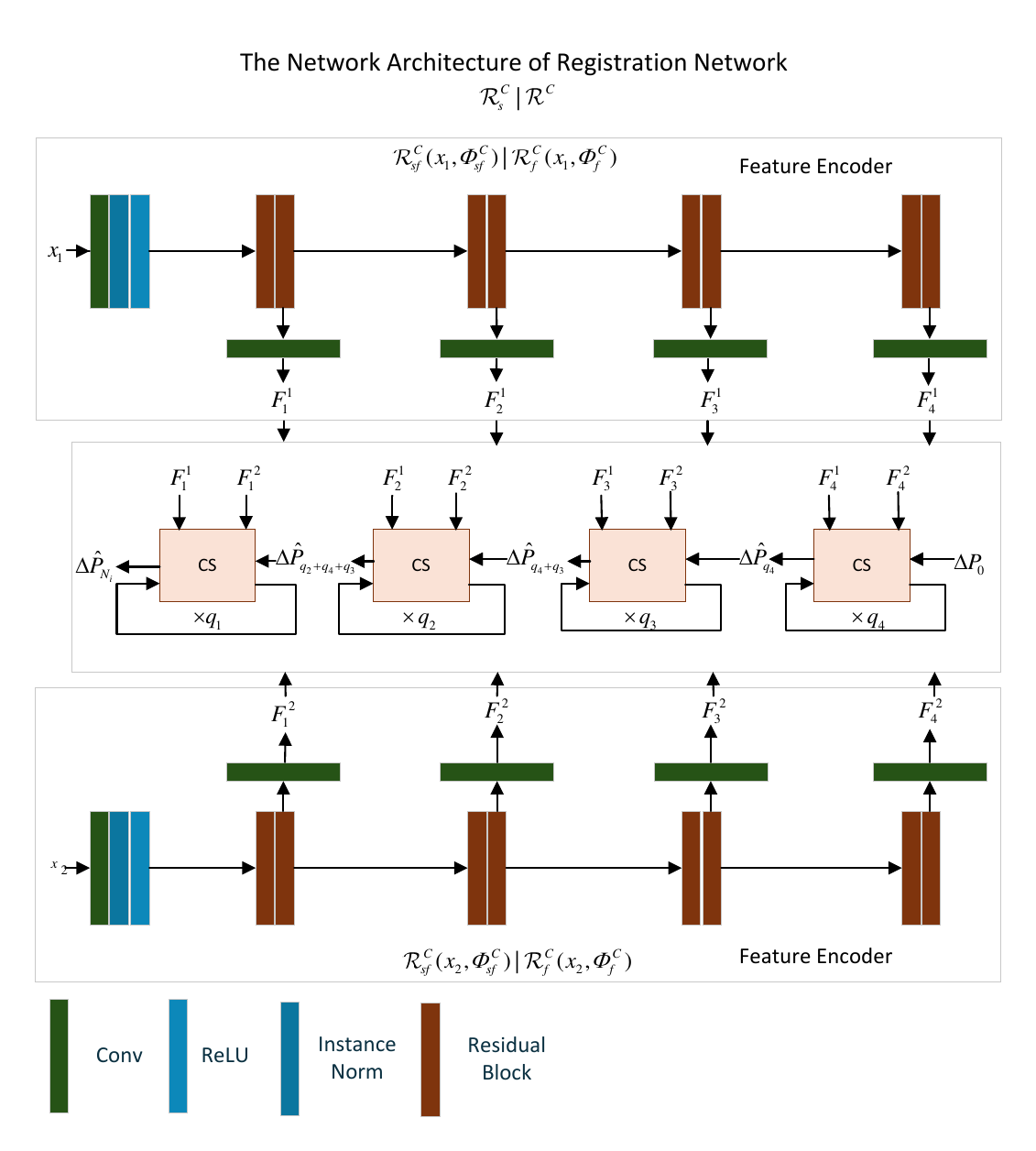}
    \caption{The network architecture of Intermediate Registration Network $\mathcal{R}^C_s$ and registration network $\mathcal{R}^C$, which is MCNet\cite{zhu2024mcnet} with four scales. The $\vec{x}_1$ and $\vec{x}_2$ are input images, which are firstly input into shared feature encoder $\mathcal{R}^C_{sf}$ or $\mathcal{R}^C_f$ to extract multiscale features $\{F_{sc}^1\}_{sc=1}^{sc=N_s}$ and $\{F_{sc}^2\}_{sc=1}^{sc=N_s}$, respectively. For MCNet in this paper, $N_s=4$. $\Delta P_0$ is the initial estimation and is set to $\{0\}^{4\times 2}$, $\{q_{sc}\}_{sc=1}^{sc=N_s}$ is the number of iterations for each scale, $\Delta \hat{P}_{N_i}$ is the final estimation, $N_i$ is the number of iterations and $\{\Delta \hat{P}_{i}\}_{i=1}^{i=N_i}$ is supervised for optimizing registration network.}
    \label{mcnet}
\end{figure}

Finally, as shown in Fig. \ref{r2}(C), the MIM encoder for source images $\Psi_{\mathrm{MIM}}^{s}$ is trained. The parameters $\varPhi_{\mathrm{mim}}^{s(it)}$ are optimized by minimizing three loss functions. 
The first one $\mathcal{L}_I^{mds}$ is to measure the difference between the translated source image $\vec{x}^{S \rightarrow T}$ and $\vec{x}^{T,W}$, which is given by:
\begin{equation}
    \begin{aligned}
        &\mathcal{L}_I^{mds} = \sum |\vec{x}^{S \rightarrow T}-\vec{x}^{T,W}|\circ \hat{M}^{TS}\\
        &\vec{x}^{S \rightarrow T} = \frac{\vec{x}_{t_2}^{T}}{\sqrt{\bar{\alpha}_{t_2}}} - \frac{\sqrt{1-\bar{\alpha}_{t_2}}}{\sqrt{\bar{\alpha}_{t_2}}} \hat{\varepsilon}^{S}_{t_2}\\
        &\hat{\varepsilon}^{S}_{t_2} = \Psi_{\mathrm{diff}}(\vec{x}_{t_2}^T,\vec{x}^T,\tilde{\vec{x}}_{\mathrm{MIM}}^{S},t_2,\varPhi_{\mathrm{dff}}^{(it)})
    \end{aligned}
\end{equation}
where $\vec{x}^{T,W}$ is the wrapping image of $\vec{x}^T$ with estimated transformation $\hat{H}^{-1}$, as shown in Fig. \ref{r2}(E), and $\hat{M}^{TS}$ serves to mask the padding pixels in $\vec{x}^{T,W}$. Since $\vec{x}^{T, W}$ is not the exact ground truth, in order to ensure that $\tilde{\vec{x}}^{S}_{\mathrm{MIM}}$ perverse geometric information, we use $\tilde{\vec{x}}^{S, F}_{\mathrm{MIM}}$ to constrain $\tilde{\vec{x}}^{S}_{\mathrm{MIM}}$, which is the MIM-like of $\vec{x}^S$ obtained by $\Psi_{\mathrm{MIM}}^{s}(\varPhi_{\mathrm{mim}}^{s(it-1)})$. For $it==1$, $\tilde{\vec{x}}^{S,F}_{\mathrm{MIM}} = \vec{x}^{S,N}_{\mathrm{MIM}}$. Specifically, the second loss $\mathcal{L}_I^{mms}$ is calculated by:
\begin{equation}
\mathcal{L}_I^{mms} = \sum |\tilde{\vec{x}}_{\mathrm{MIM}}^{S,F}-\tilde{\vec{x}}_{\mathrm{MIM}}^{S}|
\end{equation}

Therefore, the loss for optimizing $\varPhi_{\mathrm{mim}}^{s(it)}$ is:
\begin{equation}
    \mathcal{L}^{ms}_{I} = \lambda_{mds}\mathcal{L}_I^{mds} + \mathcal{L}_I^{mms}
\end{equation}

\subsubsection{Intermediate Registration Network $\mathcal{R}^C_s$}
\begin{figure}[!htbp]
    \centering
    \includegraphics[width=85mm]{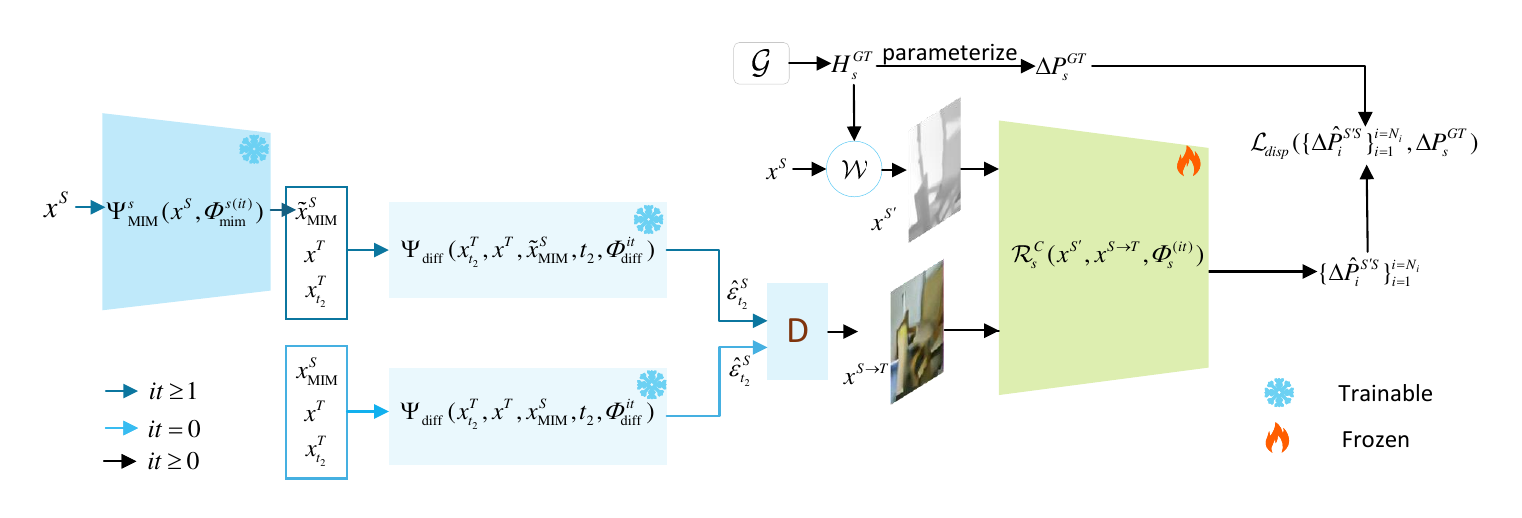}
    \caption{The training of Intermediate Registration Network $\mathcal{R}^C_s$ with cross-modality image pair $\{\vec{x}^{S^{\prime}},\vec{x}^{S\rightarrow T}\}$. $\Delta P^{GT}_s$ is the displacements of four corners by parameterizing the matrix $H^{GT}_s$. $\{\Delta \hat{P}^{S^{\prime}S}_{i}\}_{i=1}^{i=N_i}$ is the predicted displacements of corners, and $N_i$ is the number of iterations for registration network.}
    \label{l2}
\end{figure}
After training $\mathcal{T}(\varPhi_{t}^{(it)})$, our MIMGCD can translate the source image into the target domain by MIM or MIM-like feature of the source image. Since MIM or MIM-like has modality invariance, we can utilize MIM $\vec{x}_{\mathrm{MIM}}^S$ or MIM-like $\tilde{\vec{x}}_{\mathrm{MIM}}^S$ to translate the source image $x^S$ into the target domain and generate $x^{S \rightarrow T}$.
Therefore, as shown in Fig. \ref{l2}, we train intermediate registration $\mathcal{R}_{s}^C(\varPhi_{s}^{C(it)})$ with the self-supervised cross-modality image pairs $\{\vec{x}^{S^{\prime}},\vec{x}^{S\rightarrow T}\}$, where $\vec{x}^{S^{\prime}}$ is obtained by randomly wrapping $\vec{x}^S$ with transformation $H^{GT}_s$.
And for $it==0$, $\hat{\vec{x}}^{S \rightarrow T}$ is generated by reverse process $\Psi_{\mathrm{diff}}(\vec{x}_{t_2}^T,\vec{x}^T,\vec{x}_{\mathrm{MIM}}^S,t_2,\varPhi_{\mathrm{diff}}^{(0)})$ and denoising; for $it\ge 1$, we utilize the $\tilde{\vec{x}}_{\mathrm{MIM}}^S$ extracted by frozen $\Psi_{\mathrm{MIM}}^{s}(\varPhi_{\mathrm{mim}}^{s(it)})$ to generate $\hat{\vec{x}}^{S \rightarrow T}$. In this paper, as shown in Fig. \ref{mcnet}, we adopt the MCNet\cite{zhu2024mcnet} with four scales as the registration network. And, we utilize the displacement loss $\mathcal{L}{disp}$ to directly supervise the prediction of $\mathcal{R}_s^C$, thus the loss $\mathcal{L}_{S}$ for $\mathcal{R}_s^C$ is calculated by:
\begin{equation}
\begin{aligned}
    \mathcal{L}_{S} &= \mathcal{L}_{disp}(\{\Delta \hat{P}_i^{S^{\prime}S}\}_{i=1}^{N_i},\Delta P^{GT}_{s})\\
    &=\sum_{i=1}^{i=N_{i}} |\Delta \hat{P}_i^{S^{\prime}S}-\Delta P^{GT}_s|\\
    &+ \mathcal{L}_{FGO}(\Delta \hat{P}_i^{S^{\prime}S}-\Delta P^{GT}_s)
\end{aligned}
\end{equation}
where $\{\Delta \hat{P}_i^{S^{\prime}S}\}_{i=1}^{N_i}$ is the prediction of $\mathcal{R}_s^C(\varPhi_s^{C(it)})$, $\Delta P^{GT}_s$ is the groundtruth corners displacement between $\vec{x}^{S^{\prime}}$ and $\vec{x}^S$ and is obtained by parameterizing $H^{GT}_s$, 
and $\mathcal{L}_{FGO}$ is the Fine-Grained Optimization loss proposed in \cite{zhu2024mcnet}.
\begin{figure}
    \centering
    \includegraphics[width=90mm]{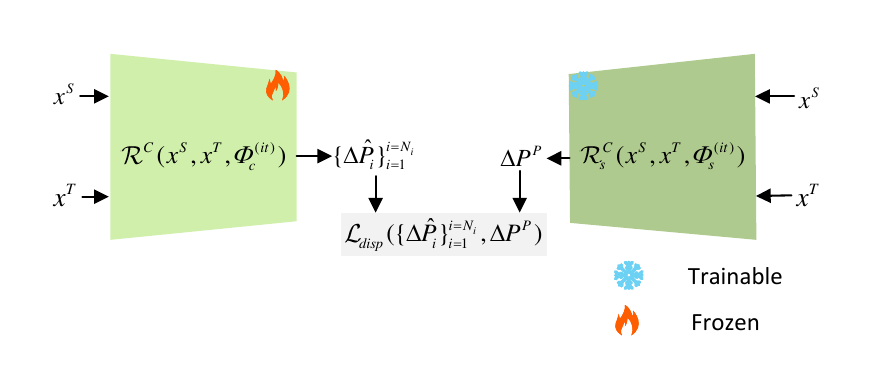}
    \caption{The traing of cross-modality registration network $\mathcal{R}^C(\varPhi_{c}^{(it)})$ with unsupervised cross-modality image pair ${x^S,x^T}$.
    $\Delta P^{Pl}$ is the pseudo-label for supervising the predicted displacements of four corners $\{\Delta \hat{P}_i\}_{i=1}^{i=N_i}$}
    \label{r1}
\end{figure}

\subsubsection{Cross-modality Registration Network $\mathcal{R}^C$}
To make the registration network more suitable for the real cross-modal image registration task, after the training of $\mathcal{R}_s^C(\varPhi_{s}^{(it)})$, as shown in Fig. \ref{r1}, we utilize the pseudo-label $\Delta P_{Pl}$ obtained by the final prediction of $R^C_s$ to supervise the output of $\mathcal{R^C}$ for unsupervised cross-modality image pair $\{\vec{x}^S,\vec{x}^T\}$. As shown in Fig. \ref{mcnet}, $R^C$ has the same architecture with $R^C_s$. Therefore, this is equivalent to first training on large synthetic data with precision labels, and then fine-tuning on real data with only noisy labels.

 the loss for training $R^C_s$ is given by:
 \begin{equation}
 \begin{aligned}
     &\mathcal{L}_{U} = \mathcal{L}_{disp}(\{\Delta\hat{P}_{i}\}_{i=1}^{i=N_i},\Delta P_{P})\\
     &\{\Delta\hat{P}_{i}\}_{i=1}^{i=N_i} = \mathcal{R}^C(\vec{x}^S,\vec{x}^T,\varPhi_c^{(it)})\\
     &\Delta P^{P} = \Delta \hat{P}^{ST}_{N_i}\\
     & \{\Delta \hat{P}^{ST}_i\}_{i=1}^{i=N_i} = \mathcal{R}_s^C(\vec{x}^S,\vec{x}^T,\varPhi_{s}^{(it)})\\
 \end{aligned}
 \end{equation}
  \begin{algorithm}[!htbp]
\caption{The training for CoLReg}
\begin{algorithmic}[1]
 \label{a1}
\Require The training hyperparameters for each network; The number of times for alternating optimization: $\mathrm{IT}$; 
The multimodal trainset with $\mathrm{N}$ unaligned image pairs. 
\Ensure $\mathcal{R}^C(\varPhi_c^{(IT-1)})$
\For{$it$ to $\mathrm{IT}-1$}
    \If{$it$ == 0}
    \State Training $\Psi_{\mathrm{diff}}(\varPhi_{\mathrm{diff}}^{(0)})$ with $\mathrm{iter_{diff}^{(0)}}$ iterations
    \State Training $\mathcal{R}_s^C(\varPhi_s^{(0)})$ with $\mathrm{iter_{s}^{(0)}}$ iterations
    \State Loading $\mathcal{R}^C$ with pretrained parameters $\varPhi_s^{(0)}$
    \State Training $\mathcal{R}^C(\varPhi_{c}^{(0)})$ with $\mathrm{iter_{c}^{(0)}}$ iterations
    \Else
    \State Loading $\Psi_{\mathrm{MIM}}^T$ with $\varPhi_{\mathrm{mim}}^{s(it-1)}$($it>=2$)
    \State Training $\Psi_{\mathrm{MIM}}^T(\varPhi_{\mathrm{mim}}^{t(it)})$ with $\mathrm{iter_{mim}^{t(it)}}$ iterations
    \State Loading $\Psi_{\mathrm{diff}}$ with $\varPhi_{\mathrm{diff}}^{(it-1)}$
    \State Training $\Psi_{\mathrm{diff}}(\varPhi_{\mathrm{diff}}^{(it)})$ with $\mathrm{iter_{diff}^{(it)}}$ iterations
    \State Loading $\Psi_{\mathrm{MIM}}^S$ with $\varPhi_{\mathrm{mim}}^{t(it)}$
    \State Training $\Psi_{\mathrm{MIM}}^S(\varPhi_{\mathrm{mim}}^{s(it)})$ with $\mathrm{iter_{mim}^{s(it)}}$ iterations 
    \State Loading $\mathcal{R}_s^C$ with $\varPhi_c^{(it-1)}$
    \State Training $\mathcal{R}_s^C(\varPhi_s^{(it)})$ with $\mathrm{iter_{s}^{(it)}}$ iterations
    \State Loading $\mathcal{R}^C$ with $\varPhi_s^{(it)}$
    \State Training $\mathcal{R}^C(\varPhi_{c}^{(it)})$ with $\mathrm{iter_{c}^{(it)}}$ iterations
    \EndIf
\EndFor
\end{algorithmic}
\end{algorithm}

 \subsubsection{The Detail of Training and Testing}
Alg.~\ref{a1} introduces the details of training CoLReg. In the initial stage($it==0$), to ensure the stability of training, we only train the conditional diffusion model  $\Psi_{\mathrm{diff}}(\varPhi_{\mathrm{diff}}^{(0)})$ of the image-to-image translation network $\mathcal{T}$ using a self-supervised dataset with perfect ground-truth. And in order to speed up the convergence of the cross-modal registration network $\mathcal{R}^C$ and improve its generalization, we use
the parameters $\varPhi_s^{(0)}$ trained by synthetic self-supervised cross-modality data to initialize $\mathcal{R}^C$. In refinement stage($it>=1$), for each 
time of alternating optimization, we utilize the pretrained parameters $\varPhi_{\mathrm{mim}}^{s(it-1)}$ and $\varPhi_{\mathrm{mim}}^{t(it)}$ to update $\Psi_{\mathrm{MIM}}^T$ and $\Psi_{\mathrm{MIM}}^S$, respectively, which not only helps promote the consistency of the extracted MIM-like features of cross-modality images. And we also utilize the parameters $\varPhi_c^{(it-1)}$ and $\varPhi_s^{(it)}$ to update $\mathcal{R}_s^C$ and $\mathcal{R}^C$, which not only makes $\mathcal{R}^C_s$ have a better initial value at the beginning of each alternating optimization, but it can also improve the generalization of $\mathcal{R}^C$.

After the training, we obtained a cross-modality registration network $\mathcal{R}^C(\varPhi_c^{(IT-1)})$. As shown in Alg.~ \ref{a2}, for testing, we can indirectly utilize  $\mathcal{R}^C$ to estimate the transformation matrix $\hat{H}$.
\begin{algorithm}[!htbp]
\caption{The inference for a cross-modality image pair}
\begin{algorithmic}[2]
\label{a2}
\Require {A cross-modality image pair: $\{\vec{x}^S,\vec{x}^T\}$;
\Require Pretrained registration network $\mathcal{R}^C(\varPhi_c^{(IT-1)})$}
\Ensure Transformation matrix $\hat{H}$ for aligning $\vec{x}^S$ with $\vec{x}^T$
\State $\Delta \hat{P}_{N_i}^{ST} = \mathcal{R}_s^C(\vec{x}^S,\vec{x}^T,\varPhi_c^{(it)})$
\State Calculating transformation matrix $\hat{H}$ according estimated displacement of four corners $\Delta \hat{P}_{N_i}^{ST}$
\end{algorithmic}
\end{algorithm}
\section{Experiment and Results}
\subsection{Experiment Setting}
\subsubsection{Dataset}
We select five open-source multimodal image datasets for our experiments, detailed as follows:

\paragraph{\textbf{GoogleEarth}~\cite{zhao2021deep}:}  
GoogleEarth contains 8,750 and 850 unaligned cross-modality image pairs in the training and testing sets, respectively.

\paragraph{\textbf{RGB\_IR\_AI}:}  
This dataset is derived from ~\cite{RAZAKARIVONY2016187}, which contains 1210 $512\times 512$ Infrared-RGB image pairs. We select 1089 pairs for training and 121 for testing. Following the perturbation method in~\cite{detone2016deep}, we finally generate 9801 unaligend cross-modality image pairs in the trainset and 2089 in the testset. 

\paragraph{\textbf{Depth\_VIS}:}  
This dataset is based on ~\cite{cho2021diml}, which provides aligned RGB and depth images across 18 indoor scenes (e.g., Warehouse, Cafe, Library). As there is a scene overlap in the original split, we redefine it by selecting 12 scenes for training and 6 for testing. We apply 10× oversampling to the original dataset. Using the same sample generation strategy as RGB\_IR\_AI, we obtain 16,210 and 4,910 unaligned image pairs for training and testing, respectively.
% This dataset is based on ~\cite{cho2021diml}, which provides aligned RGB and depth images across 18 indoor scenes (e.g., Warehouse, Cafe, Library). We select 12 scenes for training and 6 for testing. By applying geometric transformation, we obtain 16,210 and 4,910 unaligned image pairs in the trainset and testset, respectively.
\paragraph{\textbf{SAR\_Opt\_OS}:}
Based on the dataset from~\cite{xiang2020automatic}, which includes 2,011 training and 424 testing aligned SAR-optical image pairs of size $512 \times 512$, we resize all images to $192 \times 192$ and apply 5× oversampling. This yields 10,055 training pairs and 2,120 testing pairs, following the original train-test partition.
% This dataset is constructed from ~\cite{xiang2020automatic}. We utilize the aligned SAR-Optical pairs to generate the 10,055 unaligned image pairs for training and 2120 for testing.
% training, 1,190 validation, and 2,120 testing pairs.

\paragraph{\textbf{VIS\_IR\_LowLight}:}  
Sourced from LLVIP~\cite{jia2021llvip}, this dataset includes low-light RGB-infrared pairs captured at dusk across 26 street locations. Due to significant overlap between locations 25 (training) and 26 (testing), we exclude both. After resizing the original $1280 \times 1024$ images to $240 \times 192$, and following the above generation method, we obtain 11,602 unaligned training and 2,930 testing image pairs of size $128 \times 128$.

\subsubsection{Compared Methods}
We compare our MGCD-MUReg with other state-of-the-art multimodal deep learning image registration methods, as shown in Table~\ref{methods}. SPSG refers to a combination of SuperPoint~\cite{detone2018superpoint} and SuperGlue~\cite{sarlin2020superglue}, pretrained on a large supervised dataset and fine-tuned on the test set. IHN1 denote the single-scale versions of IHN. Alto, SSHNet, and SSHNet-D correspond to IHN-1 + AltO, SSHNet-IHN, and SSHNet-IHN-D proposed in~\cite{song2024unsupervised} and~\cite{yu2024internet}, respectively. We choose these configurations for comparison because, as reported in ~\cite{song2024unsupervised} and~\cite{yu2024internet}, they usually achieve better performance than combinations of Alto or SSHNet with other registration networks.
\begin{table}[!htbp]
\centering
\caption{The details of compared state-of-the-art multimodal deep learning image registration methods.}
\begin{tabular}{lll}
\toprule
Method & Year & Publication \\ 
\midrule
\rowcolor{method3} SPSG     &2020              &CVPR \\
\rowcolor{method0} DHN~\cite{detone2016deep}  & 2016 & \\
\rowcolor{method0} MHN~\cite{le2020deep}  &2020 & CVPR \\
\rowcolor{method0} IHN1~\cite{cao2022iterative} & 2022 & CVPR \\
\rowcolor{method0} IHN~\cite{cao2022iterative} & 2022 & CVPR \\
\rowcolor{method0} ReDFeat~\cite{Deng2022ReDFeatRD} & 2022 & TIP \\
\rowcolor{method0} MCNet~\cite{zhu2024mcnet} & 2024 & CVPR \\
\midrule
\rowcolor{method1} CUDHN~\cite{zhang2020content} & 2020   &ECCV\\
\rowcolor{method1} SCPNet~\cite{zhang2025scpnet} & 2024 & ECCV \\
\rowcolor{method1} Alto~\cite{song2024unsupervised} & 2024 & NIPS \\
\rowcolor{method1} SSHNet~\cite{yu2024internet} & 2025 & CVPR \\
\rowcolor{method1} SSHNet-D~\cite{yu2024internet} & 2025 & CVPR \\
\bottomrule
\end{tabular}
\begin{threeparttable}
\begin{tablenotes}
\footnotesize
\item  
\colorbox{method3}{\textbf{\textcolor{black}{Large dataset pretrained and fine-tuning on testset}}};\\
\colorbox{method0}{\textbf{\textcolor{black}{Deep supervised methods}}};\\
\colorbox{method1}{\textbf{\textcolor{black}{Deep unsupervised methods}}};
% \colorbox{method2}{\textbf{\textcolor{black}{Our unsupervised method CoDiMIR}}};\\
\end{tablenotes}
\end{threeparttable}
\label{methods}
\end{table}

\begin{table*}[!htbp]
\centering
\caption{Comparative results of different methods on five cross-modal image matching datasets (GoogleEarth, RGB\_IR\_AI, Depth\_VIS, SAR\_Opt\_OS, and VIS\_IR\_LowLight). Evaluation metrics include AUC@K (K=3,5,10,20) and MACE (lower is better). The rightmost columns report inference time and memory consumption. \textbf{Bold}: best performance.}
\resizebox{\textwidth}{!}{%
\label{table_comp}
\begin{tabular}{l|ccccc|ccccc}
\hline
Method & \multicolumn{5}{c|}{GoogleEarth} & \multicolumn{5}{c}{RGB\_IR\_AI} \\
                         &AUC@3   &AUC@5 &AUC@10 &AUC@20  &MACE   &AUC@3  &AUC@5  &AUC@10 &AUC@20  &MACE \\\hline
\rowcolor{method3}  SPSG &38.79   &59.17 &78.42  & 88.97  &2.23   &61.31 & 73.14 & 83.98 & 90.03 &5.42 \\\hline
\rowcolor{method0}  DHN  &0.01    &0.10  & 6.25  & 36.88  &12.83  &0.01   &0.39  &8.29 &37.49  &12.84 \\
\rowcolor{method0}  MHN  &1.61    &12.47  & 46.34 & 72.45  &5.51  &1.27   &7.84  &31.70 &60.84  &7.99  \\
\rowcolor{method0}  ReDFeat  &47.37 & 65.76 & 82.29 & 90.97 &1.85  & 80.56 & 87.64 & 93.17 & 96.12 &1.22  \\
\rowcolor{method0}  IHN1    &54.73 & 71.14 & 84.85 & 92.40 &1.52  & 82.23 & 88.63 & 93.51 & 96.16 &0.85 \\
\rowcolor{method0}  IHN      &63.26 & 76.68 & 87.83 &93.84  &1.23  & 90.28 &93.37 &\textbf{95.80} &\textbf{97.28} &\textbf{0.61}\\
% \rowcolor{method0}  RHWF-1   &41.74 & 60.40 & 78.51 & 89.08 &2.18  & 77.38 & 84.20 &90.16 & 94.03 &1.30  \\
% \rowcolor{method0}  RHWF-2   &65.42 & 77.64 & 88.17 & 93.97 &1.21  & 91.37 & 93.27 & 95.11 & 96.67 &0.77 \\
\rowcolor{method0}  MCNet    &\textbf{81.61} &\textbf{88.59} &\textbf{94.09} &\textbf{96.97} &\textbf{0.61}  &\textbf{93.08} &\textbf{94.21} & 95.39 & 96.62 &0.77 \\\hline
\rowcolor{method1}  CUDHN    & 0.00  &0.00  &0.00    & 2.23 &23.79      &0.00  &0.00  &0.01  &1.74  &25.14  \\
\rowcolor{method1}  SCPNet   &24.61 & 45.93 & 70.01 & 84.74 &3.06      &28.34 &46.90 &66.32 &78.50 &5.99 \\
\rowcolor{method1}  Alto     &50.73 & 67.40 & 82.47 & 90.96 &1.82      &80.37 &87.36 &92.84 &95.77 &0.94  \\
\rowcolor{method1}  SSHNet   &55.61 & 70.85 & 84.51 & 92.04 &1.60      &79.05 &86.27 &92.06 &95.34 &1.02 \\
\rowcolor{method1}  SSHNet-D &44.25 & 62.76 & 79.87 & 89.71 &2.06      &67.75 &78.34 &87.34 &92.69 &1.55 \\\hline
\rowcolor{method2}  Our      &\textbf{66.29}  &\textbf{77.60}  &\textbf{7.80}  &\textbf{93.75}  &\textbf{1.25}     &\textbf{95.23} &\textbf{96.44} &\textbf{97.35} &\textbf{97.93} &\textbf{0.50} \\\hline
Method & \multicolumn{5}{c|}{Depth\_VIS} & \multicolumn{5}{c}{SAR\_Opt\_OS} \\
                         &AUC@3   &AUC@5 &AUC@10 &AUC@20  &MACE   &AUC@3  &AUC@5  &AUC@10 &AUC@20  &MACE \\\hline
\rowcolor{method3}  SPSG & 2.47 & 6.15 & 14.36 & 25.03 &92.37 & 3.51 & 10.01 & 22.47 & 35.98  & 65.96 \\\hline
\rowcolor{method0}  DHN  & 0.00 & 0.02 & 1.08  & 19.51 &17.26  &0.00 & 0.12 & 3.59 & 25.42 & 15.88\\
\rowcolor{method0}  MHN  & 0.09 & 1.19 & 14.13 & 46.80 &10.86  &0.02 & 0.19 & 6.03 & 34.65 & 13.57\\
\rowcolor{method0}  ReDFeat  &14.67 & 29.76 & 53.78 & 73.09 &6.41  &4.94 & 16.97 & 43.17 & 67.54 &6.99\\
\rowcolor{method0}  IHN1    &31.64 & 48.77 & 67.14 & 80.22 &4.24  &10.21 & 28.31 & 55.56 & 74.55 &5.34 \\
\rowcolor{method0}  IHN      &43.44 & 57.65 & 72.57 & \textbf{83.36} &\textbf{3.58}  &\textbf{12.40} & \textbf{30.83} & \textbf{56.88} &\textbf{75.68} &\textbf{5.06}\\
% \rowcolor{method0}  RHWF-1   &18.61 & 32.87 &52.53 &69.87 &6.45 &0.01 & 0.10 & 2.93 & 23.55 &16.60\\
% \rowcolor{method0}  RHWF-2   &41.08 & 51.05 &63.27 &75.35 &5.39 & -- & -- & -- & -- &-- \\
\rowcolor{method0}  MCNet    &\textbf{57.64} &\textbf{65.48} &\textbf{74.72} &83.27 &3.66  &4.58 & 15.98 & 41.56 & 65.78 &7.11\\\hline
\rowcolor{method1}  CUDHN    &0.00 &0.00 &0.00  &1.57 &27.23 & 0.00    & 0.00  &0.00    & 1.69     &15.08\\
\rowcolor{method1}  SCPNet   &0.00  &0.00  &0.05  &1.07 &56.26 &0.00  &0.05 & 0.51 & 4.58 &32.29\\
\rowcolor{method1}  Alto     &0.00  &0.00  &0.01  &1.63 &26.32 &0.00  &0.00 & 0.00 & 1.53 &26.51 \\
\rowcolor{method1}  SSHNet   &19.03 &33.83 &54.16 & 71.12 &6.29 & 2.62 & 11.59 &34.95 &59.07 &8.80\\
\rowcolor{method1}  SSHNet-D &8.74  &20.83 &41.79 & 63.00 &7.83 & 1.93 & 9.28  &31.06  &57.64 &8.77 \\\hline
\rowcolor{method2}  Our      &\textbf{51.32} &\textbf{62.07} &\textbf{73.28} &\textbf{82.40} &\textbf{3.91}  & \textbf{18.42} &\textbf{35.46} &\textbf{56.77}  &\textbf{73.71} &\textbf{5.58} \\\hline
Method & \multicolumn{5}{c|}{VIS\_IR\_LowLight}  & \multicolumn{5}{c}{Inference Time}\\
&AUC@3   &AUC@5 &AUC@10 &AUC@20  &MACE   & \multicolumn{5}{c}{Time (ms)}  \\\hline
\rowcolor{method3}  SPSG  & 1.31  &6.66 & 20.84 & 38.49  & 70.68  & \multicolumn{5}{c}{121.77} \\\hline
\rowcolor{method0}  DHN   & 0.00  &0.00 & 0.25 & 8.45 &21.71  & \multicolumn{5}{c}{13.37}  \\
\rowcolor{method0}  MHN   & 0.00  &0.03 & 1.39 & 17.62 &18.37 & \multicolumn{5}{c}{32.58} \\
\rowcolor{method0}  ReDFeat  &0.01 &0.32  &3.32  &15.65 &135.24 & \multicolumn{5}{c}{125.44}  \\
\rowcolor{method0}  IHN1    &0.15 & 1.51 & 9.29 & 29.40 &16.56  & \multicolumn{5}{c}{90.20}  \\
\rowcolor{method0}  IHN    &0.25 & 2.37 & 12.30 & 33.02 &15.52 & \multicolumn{5}{c}{139.24}\\
% \rowcolor{method0}  RHWF-1   &-- & -- & -- & -- & --  & \multicolumn{5}{c}{171.44} \\
% \rowcolor{method0}  RHWF-2   &-- & -- & -- & -- & -- & \multicolumn{5}{c}{354.04}  \\
\rowcolor{method0}  MCNet     &\textbf{1.25} &\textbf{3.96} &\textbf{15.20} &\textbf{39.89} &\textbf{13.16} & \multicolumn{5}{c}{92.09}  \\\hline
\rowcolor{method1}  CUDHN     &0.00 &0.00 &0.00  &1.69 &25.08 & \multicolumn{5}{c}{73.14}   \\
\rowcolor{method1}  SCPNet    &0.00 &0.00 &0.02  &0.87 &34.08  & \multicolumn{5}{c}{24.64}  \\
\rowcolor{method1}  Alto      &0.00 &0.00 &0.03  &1.73 &26.33 & \multicolumn{5}{c}{52.02} \\
\rowcolor{method1}  SSHNet    &0.00 &0.00 &0.00  &1.62 &27.08 & \multicolumn{5}{c}{126.94}  \\
\rowcolor{method1}  SSHNet-D  &0.00 &0.00 &0.00  &0.57 &32.85 & \multicolumn{5}{c}{95.51}  \\\hline
\rowcolor{method2}  Our       &\textbf{5.90} &\textbf{15.76}  &\textbf{36.63} &\textbf{60.03} &\textbf{8.41}& \multicolumn{5}{c}{98.24} \\\hline
\end{tabular}
}
\begin{threeparttable}
\begin{tablenotes}
\footnotesize
\item  
\colorbox{method3}{\textbf{\textcolor{black}{Large dataset pretrained and fine-tuning on testset}}};
\colorbox{method0}{\textbf{\textcolor{black}{Deep supervised methods}}};
\colorbox{method1}{\textbf{\textcolor{black}{Deep unsupervised methods}}};
\colorbox{method2}{\textbf{\textcolor{black}{Our unsupervised method CoDiMIR}}};
\end{tablenotes}
\end{threeparttable}
\end{table*}

\begin{table*}[!htbp]
\centering
\caption{Compare our unsupervised methods CoDiMIR and other supervised methods DHN*, MHN*,ReDFeat*,IHN1*,IHN*,MCNet*. These supervised methods are trained with unaligned supervised cross-modality image pairs randomly generated in the training stage. Bold indicates the best performance, and underline indicates the second-best.}
\resizebox{\textwidth}{!}{%
\label{table_comp192}
\begin{tabular}{l|ccccc|ccccc}
\hline
Method & \multicolumn{5}{c|}{RGB\_IR\_AI} & \multicolumn{5}{c}{Depth\_VIS} \\
                             &AUC@3   &AUC@5 &AUC@10 &AUC@20    &MACE   &AUC@3  &AUC@5 &AUC@10 &AUC@20  &MACE \\\hline
\rowcolor{method0}  DHN*     &0.48    &5.71  &32.45 &62.94      &7.47   &0.02   &0.33   &9.10  &41.74 &11.85   \\
\rowcolor{method0}  MHN*     &10.85   &32.79 &62.08 & 80.32     &3.98   &5.04   &19.22 &47.94 &71.54  &5.75  \\
\rowcolor{method0}  ReDFeat* &74.75   &83.93 &91.12 & 94.92     &1.34   &10.14  &23.75 &47.90 &69.19  &7.40  \\
\rowcolor{method0}  IHN1*   &87.98   &92.27 &95.54 &97.30      &0.58   &43.44  &59.98 &75.49 &85.35  &3.12 \\
\rowcolor{method0}  IHN*     &94.25   &95.94 &97.24 &\textbf{97.99}     &\textbf{0.45}   &\underline{54.95}  &\underline{67.61} &\underline{79.39}&\underline{87.22} &\textbf{2.76}\\
\rowcolor{method0}  MCNet*   &\textbf{96.36}   &\textbf{97.00} &\textbf{97.53} &\textbf{97.99}      &\underline{0.50}   &\textbf{66.64}  &\textbf{73.70} &\textbf{81.00} &\textbf{87.34}  &\underline{2.79} \\\hline
\rowcolor{method2}  Our      &\underline{95.23}   &\underline{96.44} &\underline{97.35} &\underline{97.93} &\underline{0.50} &51.32  &62.07 &73.28 &82.40  &3.91 \\\hline
Method & \multicolumn{5}{c|}{SAR\_Opt\_OS} & \multicolumn{5}{c}{VIS\_IR\_LowLight} \\
                              &AUC@3   &AUC@5 &AUC@10 &AUC@20  &MACE   &AUC@3  &AUC@5  &AUC@10 &AUC@20  &MACE \\\hline
\rowcolor{method0}  DHN*      &0.41    &3.25  &21.46  &52.99  &9.56     &0.00   &0.00  &0.45   &10.16  &21.07 \\
\rowcolor{method0}  MHN*      &1.91    &12.47 &43.60  &69.71  &6.15     &0.00   &0.04  &1.28   &13.39  &20.51  \\
\rowcolor{method0}  ReDFeat*  &4.54    &15.93 &41.01  &65.40  &7.36     &0.07   &0.77  &7.08   &24.44  &43.70  \\
\rowcolor{method0}  IHN1*    &13.49   &31.81 &57.34  &75.28  &5.32     &0.19   &1.26  &7.17   &23.07  &19.08 \\
\rowcolor{method0}  IHN*      &15.62   &34.51 &\underline{59.16}  &\underline{76.05}  &\underline{5.16}     &0.10   &0.89  &6.96   &26.31  &17.04\\
\rowcolor{method0}  MCNet*    &\textbf{26.13}   &\textbf{48.66} &\textbf{71.35}  &\textbf{84.55}  &\textbf{3.20}     &\underline{4.74} &\underline{10.82} &\underline{24.59}  &\underline{49.38}  &\underline{10.64} \\\hline
\rowcolor{method2}  Our       &\underline{18.42}   &\underline{35.46} &56.77  &73.71  &5.58     &\textbf{5.90}   &\textbf{15.76} &\textbf{36.63}  &\textbf{60.03}  &\textbf{8.41} \\\hline
\end{tabular}
}
\begin{threeparttable}
\begin{tablenotes}
\footnotesize
\item  
\colorbox{method0}{\textbf{\textcolor{black}{Deep supervised methods trained}}};
\colorbox{method2}{\textbf{\textcolor{black}{Our unsupervised method CoDiMIR}}};
\end{tablenotes}
\end{threeparttable}
\end{table*}

\subsubsection{Metric}
To comprehensively evaluate the performance of our method against state-of-the-art approaches, we adopt the following metrics:

\paragraph{Average Corner Error (ACE).}
ACE quantifies geometric alignment accuracy by computing the mean Euclidean distance between the ground-truth and predicted positions of four corner points. It is defined as:
\begin{equation}
    ACE = \frac{1}{4}\sum_{i=1}^{4} \left\| H(x_i, y_i) - \hat{H}(x_i, y_i) \right\|_2
\end{equation}
where $\{(x_i, y_i)\}_{i=1}^{4}$ are the four corners of the source image, and $H$, $\hat{H}$ denote the ground-truth and estimated homography matrices, respectively. A lower ACE indicates a more accurate geometric transformation.

\paragraph{AUC@k (Area Under the Curve at threshold k).}
We compute the average corner reprojection error and evaluate the area under the cumulative error curve at four predefined pixel thresholds: 3, 5, 10, and 20. These are reported as AUC@3, AUC@5, AUC@10, and AUC@20, respectively. Each value reflects the percentage of samples whose ACE falls below the given threshold. Higher AUC@k values indicate better robustness and alignment precision.

\paragraph{Mean Average Correspondence Error (MACE).}
MACE measures the mean Euclidean distance between predicted and ground-truth keypoint correspondences across the entire dataset. It reflects the accuracy of point-wise matching. A lower MACE denotes more precise correspondence estimation.

\subsubsection{Implementation Details}
We adopt a single NVIDIA A6000 to conduct all the experiments.  $N$ is the number of cross-modality image pairs in the trainset. And we set the batchsize $B=16$ for all networks in our CoLReg. We utilize the Adam optimizer to train all our networks. For $it==0$, we set the learning rate as $2.5e-4$ to train $\Psi_{diff}$ with 600K iterations. For training $\mathcal{R}_s^C$ at $it==0$, we adopt the OneCycleLR scheduler with max learning rate $4e-4$ to train 300K iterations. For $it>=1$, $\Psi_{\mathrm{MIM}}^{S}$, $\Psi_{\mathrm{MIM}}^{T}$,  $\Psi_{\mathrm{diff}}$, $\mathcal{R}_s^C$ are trained by 30K, 30K,60K, 60K iterations, respectively. For $it>=0$, $\mathcal{R}^C$ is trained with 10K iterations.

\subsection{Compared with Deep Learning Methods}
\subsubsection{Quantitative Comparison}

\begin{table}[[!htbp]]
\caption{Ablation analysis of different unsupervised image registration by replacing the registration network in Alto, SSHNet, and SSHNet-D with the registration network used in our CoLReg.}
\centering
\resizebox{\columnwidth}{!}{%
\label{table_A1}
\begin{tabular}{l|cccccc}
\hline
Method & \multicolumn{5}{c}{GoogleEarth} \\
                                &AUC@3              &AUC@5            &AUC@10          &AUC@20    &MACE  \\\hline
Alto\textsuperscript{*}         &50.53            &66.08          &81.18         &90.21    &1.97\\ 
SHHNet\textsuperscript{*}       &65.06            &75.84          &86.22         &92.65    &1.47  \\ 
SHHNet-D\textsuperscript{*}     &54.99            &68.03          &81.87         &90.58    &1.88      \\ 
Our                             &\textbf{66.29}   &\textbf{77.60} &\textbf{87.80}&\textbf{93.75} &\textbf{1.25} \\  \hline
Method & \multicolumn{5}{c}{RGB\_IR\_AI} \\
                                 &AUC@3              &AUC@5            &AUC@10          &AUC@20    &MACE  \\\hline
Alto\textsuperscript{*}          &89.89            &92.06          &94.09         &95.67   &1.09\\ 
SHHNet\textsuperscript{*}        &94.95            &95.98          &96.93         &97.72   &0.53  \\ 
SHHNet-D\textsuperscript{*}      &80.82            &84.07          &88.25         &92.5    &1.57\\ 
Our                              &\textbf{95.23}   &\textbf{96.44} &\textbf{97.35} &\textbf{97.93}   &\textbf{0.50}\\ \hline  
Method & \multicolumn{5}{c}{Depth\_VIS} \\
                                 &AUC@3              &AUC@5            &AUC@10          &AUC@20    &MACE  \\\hline
Alto\textsuperscript{*}          &0.00             &0.00           &0.01          &1.42    &26.68        \\ 
SHHNet\textsuperscript{*}        &27.81            &42.29          &59.86         &74.59   &5.41  \\ 
SHHNet-D\textsuperscript{*}      &27.64            &40.54          &57.67         &73.44   &5.57      \\  
Our                              &\textbf{51.32}   &\textbf{62.07} &\textbf{73.28} &\textbf{82.40}   &\textbf{3.91}  \\  \hline  
Method & \multicolumn{5}{c}{SAR\_Opt\_OS} \\
                              &AUC@3            &AUC@5          &AUC@10        &AUC@20  &MACE  \\\hline
Alto\textsuperscript{*}       &0.00             &0.00           &0.01          &1.47    &26.23\\ 
SHHNet\textsuperscript{*}     &0.74             &4.60           &19.94         &44.80   &12.10   \\ 
SHHNet-D\textsuperscript{*}   &0.69             &4.42           &22.63         &51.77   &10.07   \\ 
Our                           & \textbf{18.42} &\textbf{35.46} &\textbf{56.77}  &\textbf{73.71} &\textbf{5.58}  \\  \hline
Method & \multicolumn{5}{c}{VIS\_IR\_LowLight} \\
                              &AUC@3              &AUC@5            &AUC@10          &AUC@20    &MACE  \\\hline
Alto\textsuperscript{*}       &0.00             &0.00           &0.01          &1.65    &26.35\\ 
SHHNet\textsuperscript{*}     &0.54             &2.47           &9.77          &27.74   &16.81 \\ 
SHHNet-D\textsuperscript{*}   &0.25             &1.30           &7.13          &26.40   &16.45   \\ 
Our                           &\textbf{5.90} &\textbf{15.76}  &\textbf{36.63} &\textbf{60.03} &\textbf{8.41} \\  \hline
\end{tabular}}
\end{table}

\begin{table}[!htbp]
\centering
\caption{Ablation analysis of image-to-image translation network MIMGCD with other unpaired multimodal image-to-image networks on RGB\_IR\_AI dataset, when $\mathrm{IT}==1$}
\resizebox{\columnwidth}{!}{%
\label{table_a2}
\begin{tabular}{l|cccccc}
\hline
Method            &AUC@3             &AUC@5                &AUC@10               &AUC@20             &MACE  \\\hline
MUNIT             &1.86              &16.24                &47.30                &66.88              &7.49  \\ 
UGATIT            &37.63             &49.84                &60.76                &69.04              &8.33  \\ 
CUT               &82.57             &86.46                &89.75                &92.64              &1.69  \\ 
UNSB              &81.23             &83.34                &85.38                &87.36              &3.64  \\ 
Our               &\textbf{93.74}    &\textbf{95.48}       &\textbf{96.61}       &\textbf{97.78}     &\textbf{0.55}  \\ \hline
\end{tabular}}
\end{table}
\begin{figure*}[!htbp]
    \centering
    \includegraphics[width=170mm]{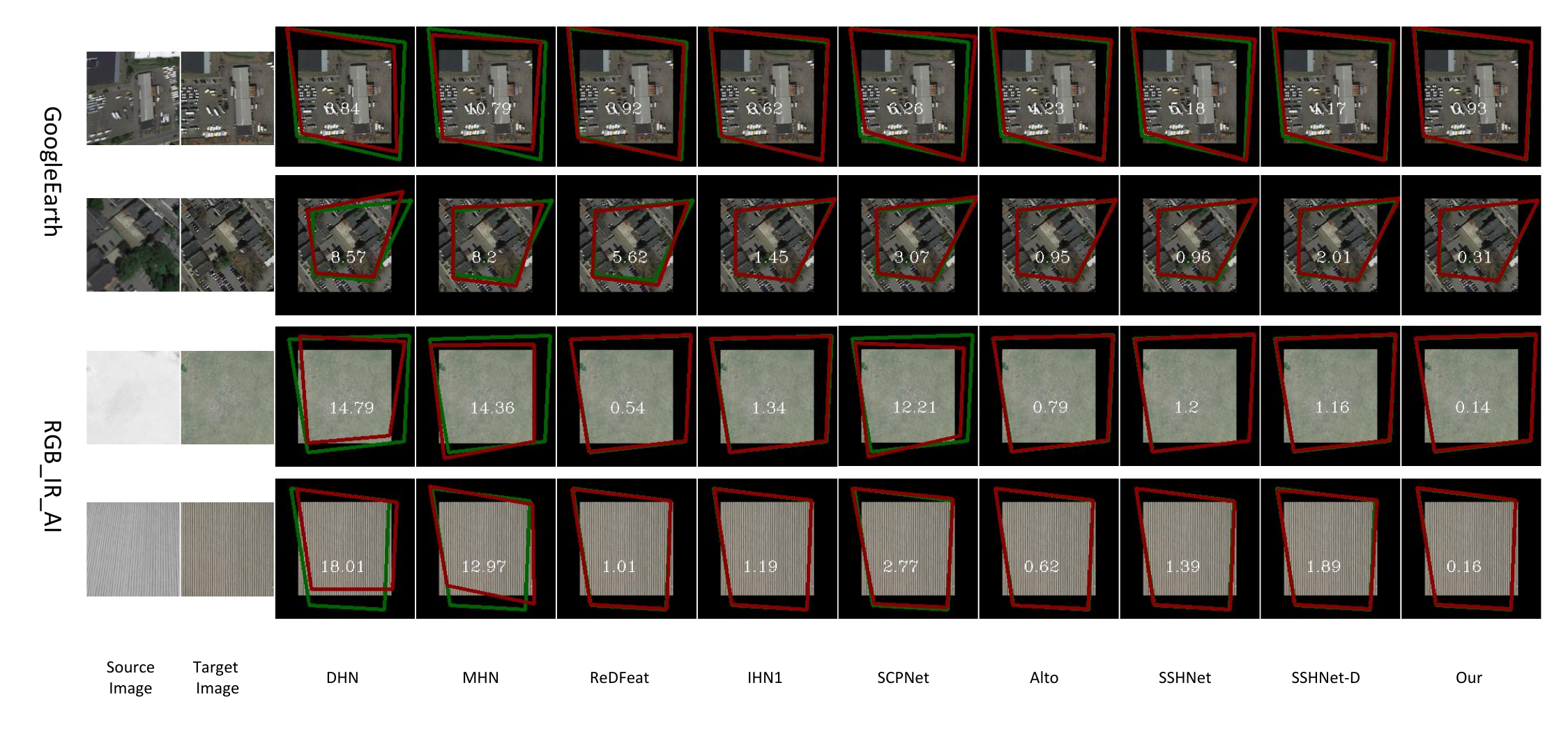}
    \caption{Qualitative image registration results of our method MGCD-UReg, supervised methods: DHN, MHM, ReDFeat, IHN1, and unsupervised methods: SCPNet, Alto, SSHNet, SSHNet-D on easy dataset GoogleEarth and RGB\_IR\_AI. Green polygons denote the ground-truth deformation from the source image to the target image. Red
polygons denote the estimated deformation using different methods on the target images. The white number denotes the average corner error.}
    \label{result_easy}
\end{figure*}

\begin{figure*}[!htbp]
    \centering
    \includegraphics[width=170mm]{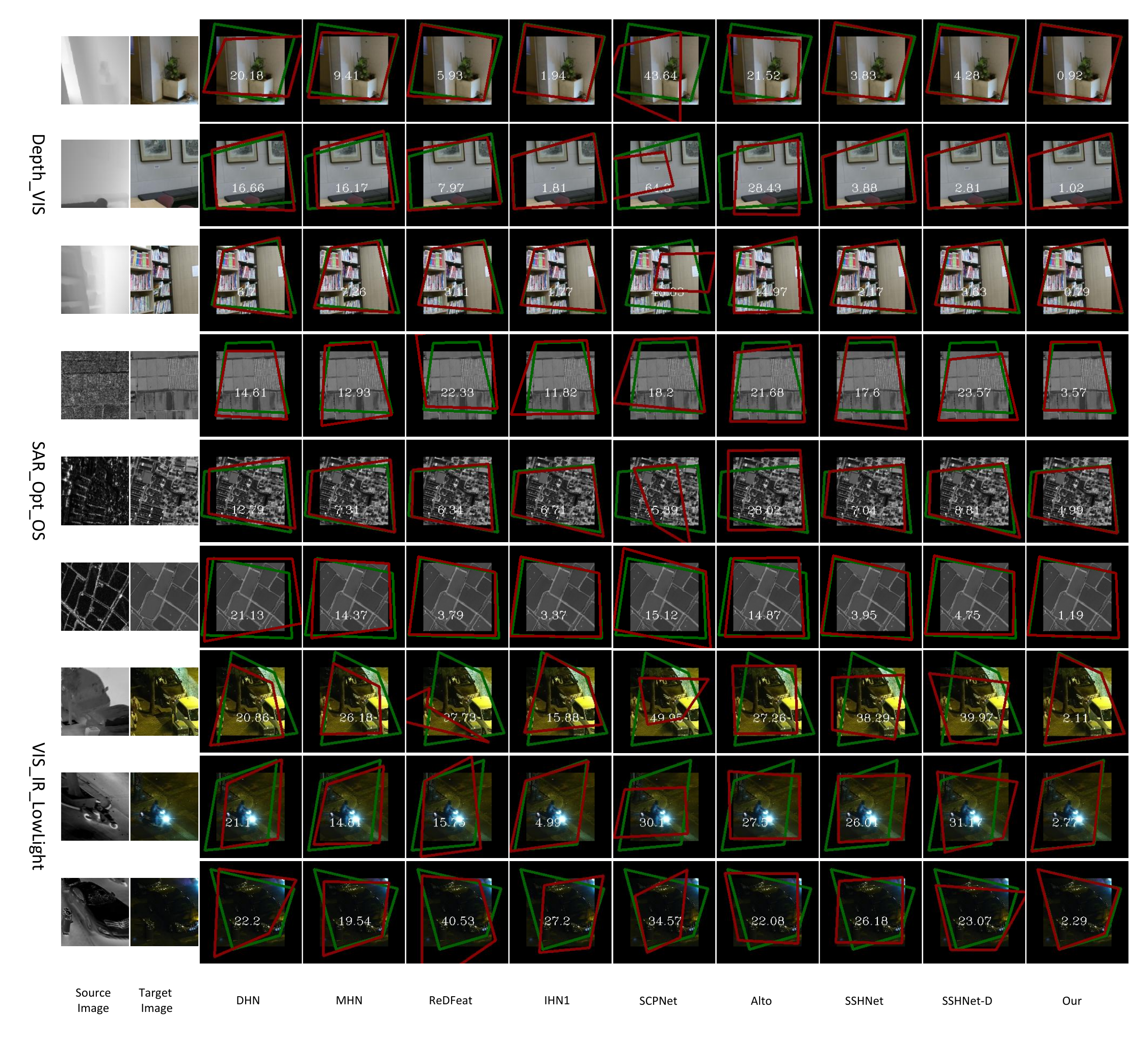}
    \caption{Qualitative image registration results of our method MGCD-UReg, supervised methods: DHN, MHM, IHN1, and unsupervised methods:  SCPNet, Alto, SSHNet, SSHNet-D on the challenge dataset. Green polygons denote the ground-truth deformation from source images to target images. Red
polygons denote the estimated deformation using different methods on the target images. The white number denotes the average corner error.}
    \label{r_gm}
\end{figure*}
\begin{figure*}[!htbp]
    \centering
    \includegraphics[width=170mm]{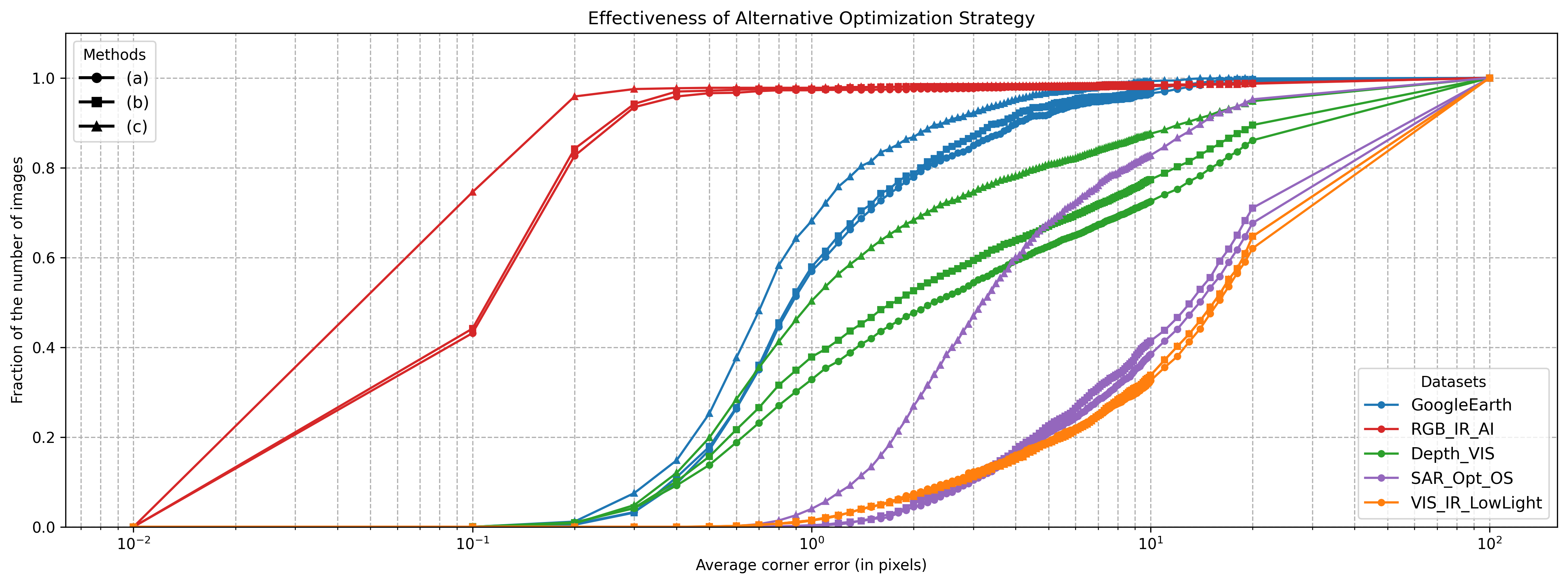}
    \caption{(a): CoLReg trained with self-supervised data and without alternative optimization($\mathrm{IT}==1$); (b): CoLReg trained with 
    unsupervised data and $\mathrm{IT}==1$; (c) CoLReg trained with unsupervised data and with alternative optimization.}
    \label{A3}
\end{figure*}
\begin{figure*}[!htbp]
    \centering
    \label{zero-shot}
    \includegraphics[width=170mm]{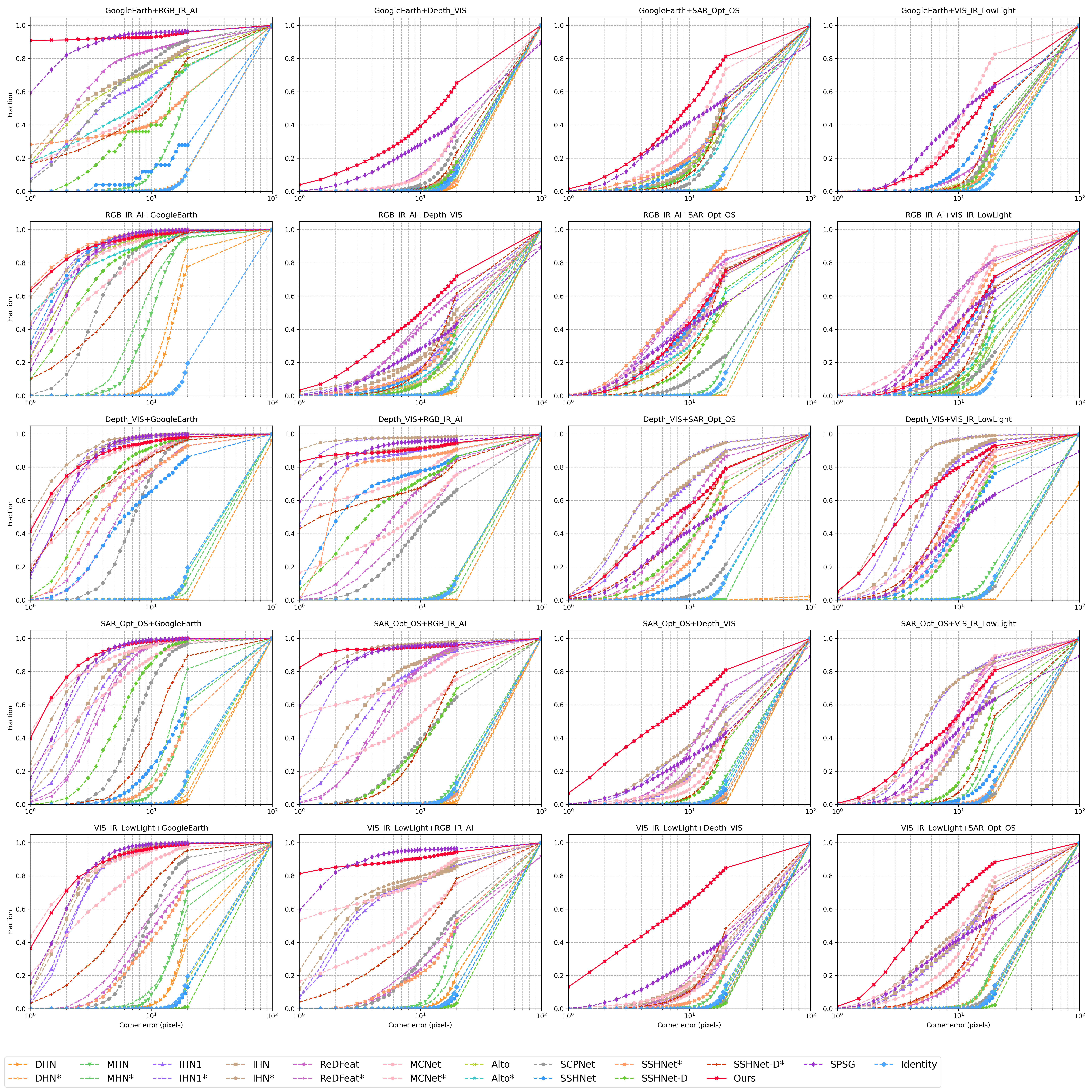}
    \caption{Compare the zero-shot image registration performance of different methods using "GoogleEarth+RGB\_IR\_AI" as an example, where GoogleEarth serves as the training set and RGB\_IR\_AI as the test set.}
\end{figure*}
For unsupervised multimodal image registration, Table.~\ref{table_comp} shows that CoLReg consistently outperforms existing unsupervised methods. Notably, it achieves significant improvements over state-of-the-art approaches SSHNet~\cite{yu2024internet} and SSHNet-D, on challenging datasets such as Depth\_VIS, SAR\_Opt\_OS, and VIS\_IR\_LowLight.

CUDHN~\cite{zhang2020content}, designed mainly for minor illumination changes, struggles with large spatial shifts and modality discrepancies, resulting in poor performance across all datasets. On easier datasets with small modality gaps (e.g., GoogleEarth, RGB\_IR\_AI), SCPNet~\cite{zhang2025scpnet} and Alto~\cite{song2024unsupervised} perform reasonably. However, on harder datasets like Depth\_VIS, SAR\_Opt\_OS, and VIS\_IR\_LowLight, SCPNet and Alto fail due to unsupervised losses being insufficient to model large geometric transformations under severe modality differences.

SSHNet uses an image-to-image (I2I) translation network to reduce modality gaps and performs better than SCPNet and Alto on some challenging datasets. Yet, its effectiveness depends heavily on the quality of I2I translation. If residual modality differences persist, the registration network, trained with mono-modal supervision, cannot handle them effectively. SSHNet-D improves generalization by distilling knowledge into a student model, but its reliance on SSHNet’s pseudo-labels still leads to performance degradation.

In contrast, our method uses MIMGCD to generate cross-modal self-supervised data, avoiding over-reliance on I2I translation. Even when the generated pairs differ from real modalities, they still guide the registration network in learning modality-invariant features. Moreover, MIMGCD better preserves geometric structures, enabling precise cross-modal registration.

Additionally, our CoLReg achieves performance comparable to state-of-the-art supervised deep learning methods.  
Our method consistently outperforms the supervised methods DHN, MHN, and ReDFeat across all datasets and evaluation metrics.
For easy dataset GoogleEarth and RGB\_IR\_AI, our method achieves performance comparable to, or even surpassing, the supervised methods IHN and MCNet. For the challenge dataset, despite the absence of precise displacement supervision, our method achieves performance comparable to state-of-the-art supervised approaches. In particular, for the high-precision registration metric AUC@3, CoLReg even surpasses most supervised methods. This advantage is pronounced on the VIS\_IR\_LowLight dataset, where limited training samples and significant lighting discrepancies lead to overfitting in supervised models. By leveraging MIMGCD to generate abundant cross-modal self-supervised data, our approach significantly enhances the network’s robustness to modality variations, resulting in superior registration performance.

In previous comparisons, all methods—both supervised and unsupervised—were trained using pre-generated unaligned cross-modality image pairs for consistency. However, for supervised methods, when aligned cross-modality image pairs are available in the training set, unaligned pairs can be dynamically generated during training by applying random geometric transformations. To ensure a fairer comparison, in Table.~\ref{table_comp192}, we compare our CoLReg with supervised methods (DHN*, MHN*, ReDFeat*, IHN1*, IHN, and MCNet*) trained using this dynamic generation strategy. Even when the advantages of supervised methods are fully leveraged, our unsupervised MGC-UMReg still outperforms the supervised methods DHN*, MHN*, ReDFeat, and IHN1* on the easy dataset RGB\_IR\_AI, and achieves performance comparable to IHN* and MCNet*. On the more challenging datasets Depth\_VIS and SAR\_Opt\_OS, our method still surpasses most supervised approaches. Notably, on the VIS\_IR\_LowLight dataset, MGC-UMReg outperforms all supervised methods and achieves the best results. These results demonstrate that CoLReg can effectively expand modality diversity during training, thereby enhancing generalization to unseen test scenarios.

\subsection{Ablation Analysis}

\subsubsection{Qualitative Comparison}
For the GoogleEarth dataset, one primary challenge arises from changes in ground objects due to differences in image acquisition times. 
These changes include natural variations, such as seasonal transitions and vegetation growth—and anthropogenic alterations like urban expansion and infrastructure development. Such variations significantly modify the image content, leading to unstable or missing feature points at the same geographic location across different time periods, thereby complicating the registration process. 
As shown in Fig.~\ref{result_easy}, we quantitatively compare our method with other approaches in regions with more pronounced changes. Our method not only outperforms state-of-the-art unsupervised methods but also surpasses supervised methods, including DHN, MHN, and IHN1. In addition to nonlinear radiation differences and geometric differences, the absence of salient objects is another crucial factor limiting image registration performance for RGB\_IR\_AI. Fig.~\ref{result_easy} compares different methods on images with low-texture areas, demonstrating the superiority of our approach over others. In challenging datasets, substantial modal discrepancies exist between source and target images. These extend beyond radiometric variations to include pronounced differences in object textures, structural details, and overall appearance. As illustrated in the Fig~.\ref{r_gm}, we conduct a qualitative comparison across three representative datasets: Depth\_VIS, SAR\_Opt\_OS, and VIS\_IR\_LowLight. On Depth\_VIS and VIS\_IR\_LowLight, our method consistently exhibits superior alignment performance across diverse scenes and spatial locations. Notably, for SAR\_Opt\_OS, where the source and target images differ significantly in imaging mechanisms and geometric properties, our method outperforms several state-of-the-art supervised deep learning approaches, highlighting its robustness under extreme modality disparities.

\subsubsection{Effectiveness of Unsupervised Multimodal Registration Framework}
 To validate the effectiveness of the unsupervised multimodal registration framework proposed in this paper, we compare the proposed CoLReg with other unsupervised registration methods that use the same registration network, as shown in Table.~ \ref{table_A1}. Specifically, we replace the registration network in Alto, SSHNet, and SSHNet-D with the MCNet used in our CoLReg, denoting the modified versions as Alto\textsuperscript{*}, SSHNet\textsuperscript{*}, and SSHNet-D\textsuperscript{*}.
As expected, the results demonstrate the effectiveness of the proposed unsupervised multimodal image registration framework.
Notably, on the particularly challenging datasets Depth\_VIS, SAR\_Opt\_OS, and VIS\_IR\_LowLight, our method consistently and substantially outperforms the current state-of-the-art unsupervised multimodal registration approaches, SSHNet\textsuperscript{*} and SSHNet-D\textsuperscript{*}, even when employing the same underlying registration network.

\subsubsection{Effectiveness of MIMGCD}
In Sec. \ref{mimgcd}, we introduce our unsupervised image-to-image translation network MIMGCD. To verify the effectiveness of MIMGCD, we compared it with four unsupervised image-to-image translation networks, MUNIT \cite{Huang2018MultimodalUI}, UGATIT \cite{Kim2020U-GAT-IT}, CUT \cite{park2020cut}, and UNSB \cite{kim2023unsb} as shown in the Tab. \ref{table_a2}. These image-to-image networks do not consider the imperfect ground-truth supervision, therefore, for fair comparison, we only show the performance of different image-to-image translation networks on the cross-modal image registration task when $\mathrm{IT}=1$, that is, without using $\vec{x}^{T, W}$ supervision. We observe that our MIMGCD outperforms other unsupervised image-to-image translation networks across various metrics. These results demonstrate that our proposed unsupervised image-to-image translation network generates high-quality cross-modal image pairs that preserve geometric features while maintaining significant modality differences.

\subsubsection{Effectiveness of Alternative Optimization}
To verify the effectiveness of our proposed alternative optimization strategy, we evaluate CoLReg under four different configurations, as illustrated in Fig. \ref{A3}. The model trained solely on self-supervised data without alternative optimization performs the worst. Introducing unsupervised data for refinement improves performance. Finally, applying both the alternative optimization strategy and training $\mathcal{R}^C$ yields the best performance. These results demonstrate that our alternative optimization effectively bridges the gap between generated self-supervised and real multimodal image pairs. Moreover, incorporating the training of $\mathcal{R}^C$ with real unsupervised image pairs further enhances the model’s suitability for cross-modality registration.

\subsection{Zero-Shot Analysis}
We build zero-shot cross-modality image registration tasks to evaluate the generalization of our method and other deep learning methods.
As shown in Fig.~\ref{zero-shot}, We train the model separately on the training sets of GoogleEarth, RGB\_IR\_AI, Depth\_VIS, SAR\_Opt\_OS, and VIS\_IR\_LowLight. Then, we evaluate its generalization performance by testing it on the test sets of the remaining unseen datasets. Overall, our method demonstrates better generalization performance than all other methods. In the easy dataset tests using GoogleEarth and RGB\_IR\_AI, the proportion of images achieving high-precision registration with our method is significantly higher than that of the other unsupervised methods. For the unseen Depth\_VIS dataset, our method consistently outperforms other approaches by a wide margin, regardless of the training set used. For the VIS\_IR\_LowLight dataset, where the training set contains limited scene diversity, our method achieves performance comparable to SPSG trained on large-scale data for the unseen easy dataset. For the more challenging datasets, Depth\_VIS and SAR\_Opt\_OS, our method significantly outperforms all other methods when models are only trained with VIS\_IR\_LowLight dataset.

\section{Conclusion}
In this paper, we propose CoLReg, a novel collaborative learning framework for unsupervised multimodal image registration. Instead of the conventional "image translation and mono-modal image registration" pipeline, we reformulate the task as a collaborative learning problem among three components for direct cross-modal image registration.  The tree components, a MIM-guided conditional diffusion model for image translation, a self-supervised intermediate network, and a distill network, are optimized in an alternating manner. Our framework enables stable and effective training without relying on ground-truth transformations. The introduction of the MIM-guided conditional diffusion model (MIMGCD), with learnable MIM features for unsupervised image-to-image translation, enhances the preservation of geometric information in the generated images, which facilitates the training of self-supervised cross-modal registration networks. Extensive experiments on multiple datasets validate the superiority of CoLReg over state-of-the-art unsupervised methods and show comparable performance to leading supervised approaches.

\bibliographystyle{elsarticle-harv}
\bibliography{bibtex}

\begin{thebibliography}{42}
\expandafter\ifx\csname natexlab\endcsname\relax\def\natexlab#1{#1}\fi
\providecommand{\url}[1]{\texttt{#1}}
\providecommand{\href}[2]{#2}
\providecommand{\path}[1]{#1}
\providecommand{\DOIprefix}{doi:}
\providecommand{\ArXivprefix}{arXiv:}
\providecommand{\URLprefix}{URL: }
\providecommand{\Pubmedprefix}{pmid:}
\providecommand{\doi}[1]{\href{http://dx.doi.org/#1}{\path{#1}}}
\providecommand{\Pubmed}[1]{\href{pmid:#1}{\path{#1}}}
\providecommand{\bibinfo}[2]{#2}
\ifx\xfnm\relax \def\xfnm[#1]{\unskip,\space#1}\fi
%Type = Inproceedings
\bibitem[{Arar et~al.(2020)Arar, Ginger, Danon, Bermano and Cohen-Or}]{arar2020unsupervised}
\bibinfo{author}{Arar, M.}, \bibinfo{author}{Ginger, Y.}, \bibinfo{author}{Danon, D.}, \bibinfo{author}{Bermano, A.H.}, \bibinfo{author}{Cohen-Or, D.}, \bibinfo{year}{2020}.
\newblock \bibinfo{title}{Unsupervised multi-modal image registration via geometry preserving image-to-image translation}, in: \bibinfo{booktitle}{Proceedings of the IEEE/CVF conference on computer vision and pattern recognition}, pp. \bibinfo{pages}{13410--13419}.
%Type = Inproceedings
\bibitem[{Cao et~al.(2022)Cao, Hu, Sheng and Shen}]{cao2022iterative}
\bibinfo{author}{Cao, S.Y.}, \bibinfo{author}{Hu, J.}, \bibinfo{author}{Sheng, Z.}, \bibinfo{author}{Shen, H.L.}, \bibinfo{year}{2022}.
\newblock \bibinfo{title}{Iterative deep homography estimation}, in: \bibinfo{booktitle}{Proceedings of the IEEE/CVF conference on computer vision and pattern recognition}, pp. \bibinfo{pages}{1879--1888}.
%Type = Inproceedings
\bibitem[{Cao et~al.(2023)Cao, Zhang, Luo, Yu, Sheng, Li and Shen}]{cao2023recurrent}
\bibinfo{author}{Cao, S.Y.}, \bibinfo{author}{Zhang, R.}, \bibinfo{author}{Luo, L.}, \bibinfo{author}{Yu, B.}, \bibinfo{author}{Sheng, Z.}, \bibinfo{author}{Li, J.}, \bibinfo{author}{Shen, H.L.}, \bibinfo{year}{2023}.
\newblock \bibinfo{title}{Recurrent homography estimation using homography-guided image warping and focus transformer}, in: \bibinfo{booktitle}{Proceedings of the IEEE/CVF Conference on Computer Vision and Pattern Recognition}, pp. \bibinfo{pages}{9833--9842}.
%Type = Article
\bibitem[{Chen et~al.(2022)Chen, Wei and Li}]{chen2022unsupervised}
\bibinfo{author}{Chen, Z.}, \bibinfo{author}{Wei, J.}, \bibinfo{author}{Li, R.}, \bibinfo{year}{2022}.
\newblock \bibinfo{title}{Unsupervised multi-modal medical image registration via discriminator-free image-to-image translation}.
\newblock \bibinfo{journal}{arXiv preprint arXiv:2204.13656} .
%Type = Article
\bibitem[{Cho et~al.(2021)Cho, Min, Kim and Sohn}]{cho2021diml}
\bibinfo{author}{Cho, J.}, \bibinfo{author}{Min, D.}, \bibinfo{author}{Kim, Y.}, \bibinfo{author}{Sohn, K.}, \bibinfo{year}{2021}.
\newblock \bibinfo{title}{Diml/cvl rgb-d dataset: 2m rgb-d images of natural indoor and outdoor scenes}.
\newblock \bibinfo{journal}{arXiv preprint arXiv:2110.11590} .
%Type = Article
\bibitem[{Cui et~al.(2022)Cui, Ma, Wan, Zhong, Luo and Xu}]{Cui2022CrossModalityIM}
\bibinfo{author}{Cui, S.}, \bibinfo{author}{Ma, A.}, \bibinfo{author}{Wan, Y.}, \bibinfo{author}{Zhong, Y.}, \bibinfo{author}{Luo, B.}, \bibinfo{author}{Xu, M.}, \bibinfo{year}{2022}.
\newblock \bibinfo{title}{Cross-modality image matching network with modality-invariant feature representation for airborne-ground thermal infrared and visible datasets}.
\newblock \bibinfo{journal}{IEEE Transactions on Geoscience and Remote Sensing} \bibinfo{volume}{60}, \bibinfo{pages}{1--14}.
\newblock \URLprefix \url{https://api.semanticscholar.org/CorpusID:241410294}.
%Type = Article
\bibitem[{Dang et~al.(2015)Dang, Luqman, Coustaty, Tran and Ogier}]{Dang2015SRIFSA}
\bibinfo{author}{Dang, Q.B.}, \bibinfo{author}{Luqman, M.M.}, \bibinfo{author}{Coustaty, M.}, \bibinfo{author}{Tran, D.C.}, \bibinfo{author}{Ogier, J.M.}, \bibinfo{year}{2015}.
\newblock \bibinfo{title}{Srif: Scale and rotation invariant features for camera-based document image retrieval}.
\newblock \bibinfo{journal}{2015 13th International Conference on Document Analysis and Recognition (ICDAR)} , \bibinfo{pages}{601--605}\URLprefix \url{https://api.semanticscholar.org/CorpusID:27040928}.
%Type = Article
\bibitem[{Deng and Ma(2022)}]{Deng2022ReDFeatRD}
\bibinfo{author}{Deng, Y.C.}, \bibinfo{author}{Ma, J.}, \bibinfo{year}{2022}.
\newblock \bibinfo{title}{Redfeat: Recoupling detection and description for multimodal feature learning}.
\newblock \bibinfo{journal}{IEEE Transactions on Image Processing} \bibinfo{volume}{32}, \bibinfo{pages}{591--602}.
\newblock \URLprefix \url{https://api.semanticscholar.org/CorpusID:248811367}.
%Type = Article
\bibitem[{DeTone et~al.(2016)DeTone, Malisiewicz and Rabinovich}]{detone2016deep}
\bibinfo{author}{DeTone, D.}, \bibinfo{author}{Malisiewicz, T.}, \bibinfo{author}{Rabinovich, A.}, \bibinfo{year}{2016}.
\newblock \bibinfo{title}{Deep image homography estimation}.
\newblock \bibinfo{journal}{arXiv preprint arXiv:1606.03798} .
%Type = Inproceedings
\bibitem[{DeTone et~al.(2018)DeTone, Malisiewicz and Rabinovich}]{detone2018superpoint}
\bibinfo{author}{DeTone, D.}, \bibinfo{author}{Malisiewicz, T.}, \bibinfo{author}{Rabinovich, A.}, \bibinfo{year}{2018}.
\newblock \bibinfo{title}{Superpoint: Self-supervised interest point detection and description}, in: \bibinfo{booktitle}{Proceedings of the IEEE conference on computer vision and pattern recognition workshops}, pp. \bibinfo{pages}{224--236}.
%Type = Inproceedings
\bibitem[{Guo(2025)}]{guo2025unsupervised}
\bibinfo{author}{Guo, M.}, \bibinfo{year}{2025}.
\newblock \bibinfo{title}{Unsupervised multi-modal medical image registration via invertible translation}, in: \bibinfo{booktitle}{European Conference on Computer Vision}, \bibinfo{organization}{Springer}. pp. \bibinfo{pages}{22--38}.
%Type = Inproceedings
\bibitem[{Huang et~al.(2018)Huang, Liu, Belongie and Kautz}]{Huang2018MultimodalUI}
\bibinfo{author}{Huang, X.}, \bibinfo{author}{Liu, M.Y.}, \bibinfo{author}{Belongie, S.J.}, \bibinfo{author}{Kautz, J.}, \bibinfo{year}{2018}.
\newblock \bibinfo{title}{Multimodal unsupervised image-to-image translation}, in: \bibinfo{booktitle}{European Conference on Computer Vision}.
\newblock \URLprefix \url{https://api.semanticscholar.org/CorpusID:4883312}.
%Type = Inproceedings
\bibitem[{Jia et~al.(2021)Jia, Zhu, Li, Tang and Zhou}]{jia2021llvip}
\bibinfo{author}{Jia, X.}, \bibinfo{author}{Zhu, C.}, \bibinfo{author}{Li, M.}, \bibinfo{author}{Tang, W.}, \bibinfo{author}{Zhou, W.}, \bibinfo{year}{2021}.
\newblock \bibinfo{title}{Llvip: A visible-infrared paired dataset for low-light vision}, in: \bibinfo{booktitle}{Proceedings of the IEEE/CVF international conference on computer vision}, pp. \bibinfo{pages}{3496--3504}.
%Type = Inproceedings
\bibitem[{Kim et~al.(2024)Kim, Kwon, Kim and Ye}]{kim2023unsb}
\bibinfo{author}{Kim, B.}, \bibinfo{author}{Kwon, G.}, \bibinfo{author}{Kim, K.}, \bibinfo{author}{Ye, J.C.}, \bibinfo{year}{2024}.
\newblock \bibinfo{title}{Unpaired image-to-image translation via neural schrödinger bridge}, in: \bibinfo{booktitle}{ICLR}.
%Type = Inproceedings
\bibitem[{Kim et~al.(2020)Kim, Kim, Kang and Lee}]{Kim2020U-GAT-IT}
\bibinfo{author}{Kim, J.}, \bibinfo{author}{Kim, M.}, \bibinfo{author}{Kang, H.}, \bibinfo{author}{Lee, K.H.}, \bibinfo{year}{2020}.
\newblock \bibinfo{title}{U-gat-it: Unsupervised generative attentional networks with adaptive layer-instance normalization for image-to-image translation}, in: \bibinfo{booktitle}{International Conference on Learning Representations}.
\newblock \URLprefix \url{https://openreview.net/forum?id=BJlZ5ySKPH}.
%Type = Article
\bibitem[{Koguciuk et~al.(2021)Koguciuk, Arani and Zonooz}]{Koguciuk2021PerceptualLF}
\bibinfo{author}{Koguciuk, D.}, \bibinfo{author}{Arani, E.}, \bibinfo{author}{Zonooz, B.}, \bibinfo{year}{2021}.
\newblock \bibinfo{title}{Perceptual loss for robust unsupervised homography estimation}.
\newblock \bibinfo{journal}{2021 IEEE/CVF Conference on Computer Vision and Pattern Recognition Workshops (CVPRW)} , \bibinfo{pages}{4269--4278}\URLprefix \url{https://api.semanticscholar.org/CorpusID:233307202}.
%Type = Inproceedings
\bibitem[{Kong et~al.(2023)Kong, Qi, Shen, Wang, Zhang, Hu and Zhou}]{kong2023indescribable}
\bibinfo{author}{Kong, L.}, \bibinfo{author}{Qi, X.S.}, \bibinfo{author}{Shen, Q.}, \bibinfo{author}{Wang, J.}, \bibinfo{author}{Zhang, J.}, \bibinfo{author}{Hu, Y.}, \bibinfo{author}{Zhou, Q.}, \bibinfo{year}{2023}.
\newblock \bibinfo{title}{Indescribable multi-modal spatial evaluator}, in: \bibinfo{booktitle}{Proceedings of the IEEE/CVF Conference on Computer Vision and Pattern Recognition}, pp. \bibinfo{pages}{9853--9862}.
%Type = Inproceedings
\bibitem[{Le et~al.(2020)Le, Liu, Zhang and Agarwala}]{le2020deep}
\bibinfo{author}{Le, H.}, \bibinfo{author}{Liu, F.}, \bibinfo{author}{Zhang, S.}, \bibinfo{author}{Agarwala, A.}, \bibinfo{year}{2020}.
\newblock \bibinfo{title}{Deep homography estimation for dynamic scenes}, in: \bibinfo{booktitle}{Proceedings of the IEEE/CVF conference on computer vision and pattern recognition}, pp. \bibinfo{pages}{7652--7661}.
%Type = Article
\bibitem[{Li et~al.(2019)Li, Hu and Ai}]{Li2019RIFTMI}
\bibinfo{author}{Li, J.}, \bibinfo{author}{Hu, Q.}, \bibinfo{author}{Ai, M.}, \bibinfo{year}{2019}.
\newblock \bibinfo{title}{Rift: Multi-modal image matching based on radiation-variation insensitive feature transform}.
\newblock \bibinfo{journal}{IEEE Transactions on Image Processing} \bibinfo{volume}{29}, \bibinfo{pages}{3296--3310}.
\newblock \URLprefix \url{https://api.semanticscholar.org/CorpusID:209463301}.
%Type = Article
\bibitem[{Li et~al.()Li, Shao, Qian and Zhang}]{lifddm}
\bibinfo{author}{Li, Y.}, \bibinfo{author}{Shao, H.C.}, \bibinfo{author}{Qian, X.}, \bibinfo{author}{Zhang, Y.}, .
\newblock \bibinfo{title}{Fddm: Unsupervised mr-to-ct translation with a frequency-decoupled diffusion model} .
%Type = Article
\bibitem[{Liu et~al.(2024)Liu, He and Zhang}]{Liu2024GRiDGR}
\bibinfo{author}{Liu, Y.}, \bibinfo{author}{He, W.}, \bibinfo{author}{Zhang, H.}, \bibinfo{year}{2024}.
\newblock \bibinfo{title}{Grid: Guided refinement for detector-free multimodal image matching}.
\newblock \bibinfo{journal}{IEEE Transactions on Image Processing} \bibinfo{volume}{33}, \bibinfo{pages}{5892--5906}.
\newblock \URLprefix \url{https://api.semanticscholar.org/CorpusID:273288273}.
%Type = Article
\bibitem[{Lv et~al.(2023)Lv, Huang, Sun, Lei, Benediktsson and Li}]{Lv2023NovelEU}
\bibinfo{author}{Lv, Z.}, \bibinfo{author}{Huang, H.}, \bibinfo{author}{Sun, W.}, \bibinfo{author}{Lei, T.}, \bibinfo{author}{Benediktsson, J.A.}, \bibinfo{author}{Li, J.}, \bibinfo{year}{2023}.
\newblock \bibinfo{title}{Novel enhanced unet for change detection using multimodal remote sensing image}.
\newblock \bibinfo{journal}{IEEE Geoscience and Remote Sensing Letters} \bibinfo{volume}{20}, \bibinfo{pages}{1--5}.
\newblock \URLprefix \url{https://api.semanticscholar.org/CorpusID:264334831}.
%Type = Inproceedings
\bibitem[{Park et~al.(2020)Park, Efros, Zhang and Zhu}]{park2020cut}
\bibinfo{author}{Park, T.}, \bibinfo{author}{Efros, A.A.}, \bibinfo{author}{Zhang, R.}, \bibinfo{author}{Zhu, J.Y.}, \bibinfo{year}{2020}.
\newblock \bibinfo{title}{Contrastive learning for unpaired image-to-image translation}, in: \bibinfo{booktitle}{European Conference on Computer Vision}.
%Type = Article
\bibitem[{Quan et~al.(2022)Quan, Wang, Gu, Lei, Yang, Wei, Hou and Jiao}]{Quan2022DeepFC}
\bibinfo{author}{Quan, D.}, \bibinfo{author}{Wang, S.}, \bibinfo{author}{Gu, Y.}, \bibinfo{author}{Lei, R.}, \bibinfo{author}{Yang, B.}, \bibinfo{author}{Wei, S.}, \bibinfo{author}{Hou, B.}, \bibinfo{author}{Jiao, L.}, \bibinfo{year}{2022}.
\newblock \bibinfo{title}{Deep feature correlation learning for multi-modal remote sensing image registration}.
\newblock \bibinfo{journal}{IEEE Transactions on Geoscience and Remote Sensing} \bibinfo{volume}{60}, \bibinfo{pages}{1--16}.
\newblock \URLprefix \url{https://api.semanticscholar.org/CorpusID:251354579}.
%Type = Article
\bibitem[{Razakarivony and Jurie(2016)}]{RAZAKARIVONY2016187}
\bibinfo{author}{Razakarivony, S.}, \bibinfo{author}{Jurie, F.}, \bibinfo{year}{2016}.
\newblock \bibinfo{title}{Vehicle detection in aerial imagery : A small target detection benchmark}.
\newblock \bibinfo{journal}{Journal of Visual Communication and Image Representation} \bibinfo{volume}{34}, \bibinfo{pages}{187--203}.
\newblock \URLprefix \url{https://www.sciencedirect.com/science/article/pii/S1047320315002187}, \DOIprefix\doi{https://doi.org/10.1016/j.jvcir.2015.11.002}.
%Type = Inproceedings
\bibitem[{Sarlin et~al.(2020)Sarlin, DeTone, Malisiewicz and Rabinovich}]{sarlin2020superglue}
\bibinfo{author}{Sarlin, P.E.}, \bibinfo{author}{DeTone, D.}, \bibinfo{author}{Malisiewicz, T.}, \bibinfo{author}{Rabinovich, A.}, \bibinfo{year}{2020}.
\newblock \bibinfo{title}{Superglue: Learning feature matching with graph neural networks}, in: \bibinfo{booktitle}{Proceedings of the IEEE/CVF conference on computer vision and pattern recognition}, pp. \bibinfo{pages}{4938--4947}.
%Type = Article
\bibitem[{Sasaki et~al.(2021)Sasaki, Willcocks and Breckon}]{Sasaki2021UNITDDPMUI}
\bibinfo{author}{Sasaki, H.}, \bibinfo{author}{Willcocks, C.G.}, \bibinfo{author}{Breckon, T.}, \bibinfo{year}{2021}.
\newblock \bibinfo{title}{Unit-ddpm: Unpaired image translation with denoising diffusion probabilistic models}.
\newblock \bibinfo{journal}{ArXiv} \bibinfo{volume}{abs/2104.05358}.
\newblock \URLprefix \url{https://api.semanticscholar.org/CorpusID:233210328}.
%Type = Article
\bibitem[{Song et~al.(2024)Song, Lew, Jang and Yoon}]{song2024unsupervised}
\bibinfo{author}{Song, S.}, \bibinfo{author}{Lew, J.}, \bibinfo{author}{Jang, H.}, \bibinfo{author}{Yoon, S.}, \bibinfo{year}{2024}.
\newblock \bibinfo{title}{Unsupervised homography estimation on multimodal image pair via alternating optimization}.
\newblock \bibinfo{journal}{arXiv preprint arXiv:2411.13036} .
%Type = Article
\bibitem[{Wei et~al.(2025)Wei, Guo, Yu, Wei and Li}]{wei2025osdm}
\bibinfo{author}{Wei, X.}, \bibinfo{author}{Guo, W.}, \bibinfo{author}{Yu, W.}, \bibinfo{author}{Wei, F.}, \bibinfo{author}{Li, D.}, \bibinfo{year}{2025}.
\newblock \bibinfo{title}{Osdm-mreg: Multimodal image registration based one step diffusion model}.
\newblock \bibinfo{journal}{arXiv preprint arXiv:2504.06027} .
%Type = Article
\bibitem[{Xiang et~al.(2020)Xiang, Tao, Wang, You and Han}]{xiang2020automatic}
\bibinfo{author}{Xiang, Y.}, \bibinfo{author}{Tao, R.}, \bibinfo{author}{Wang, F.}, \bibinfo{author}{You, H.}, \bibinfo{author}{Han, B.}, \bibinfo{year}{2020}.
\newblock \bibinfo{title}{Automatic registration of optical and sar images via improved phase congruency model}.
\newblock \bibinfo{journal}{IEEE Journal of Selected Topics in Applied Earth Observations and Remote Sensing} \bibinfo{volume}{13}, \bibinfo{pages}{5847--5861}.
%Type = Article
\bibitem[{Xiao et~al.(2024a)Xiao, Pu, Chen and Gao}]{Xiao2024DGFNetDC}
\bibinfo{author}{Xiao, F.}, \bibinfo{author}{Pu, Z.}, \bibinfo{author}{Chen, J.}, \bibinfo{author}{Gao, X.}, \bibinfo{year}{2024}a.
\newblock \bibinfo{title}{Dgfnet: Depth-guided cross-modality fusion network for rgb-d salient object detection}.
\newblock \bibinfo{journal}{IEEE Transactions on Multimedia} \bibinfo{volume}{26}, \bibinfo{pages}{2648--2658}.
\newblock \URLprefix \url{https://api.semanticscholar.org/CorpusID:260592945}.
%Type = Article
\bibitem[{Xiao et~al.(2024b)Xiao, Zhang, Tortei and Loianno}]{Xiao2024STHNDH}
\bibinfo{author}{Xiao, J.}, \bibinfo{author}{Zhang, N.}, \bibinfo{author}{Tortei, D.}, \bibinfo{author}{Loianno, G.}, \bibinfo{year}{2024}b.
\newblock \bibinfo{title}{Sthn: Deep homography estimation for uav thermal geo-localization with satellite imagery}.
\newblock \bibinfo{journal}{IEEE Robotics and Automation Letters} \bibinfo{volume}{9}, \bibinfo{pages}{8754--8761}.
\newblock \URLprefix \url{https://api.semanticscholar.org/CorpusID:270199523}.
%Type = Article
\bibitem[{Xu et~al.(2022)Xu, Ma, Yuan, Le and Liu}]{Xu2022RFNetUN}
\bibinfo{author}{Xu, H.}, \bibinfo{author}{Ma, J.}, \bibinfo{author}{Yuan, J.}, \bibinfo{author}{Le, Z.}, \bibinfo{author}{Liu, W.}, \bibinfo{year}{2022}.
\newblock \bibinfo{title}{Rfnet: Unsupervised network for mutually reinforcing multi-modal image registration and fusion}.
\newblock \bibinfo{journal}{2022 IEEE/CVF Conference on Computer Vision and Pattern Recognition (CVPR)} , \bibinfo{pages}{19647--19656}\URLprefix \url{https://api.semanticscholar.org/CorpusID:250602199}.
%Type = Article
\bibitem[{Ye et~al.(2022)Ye, Tang, Zhu, Yang, Li and Hao}]{9758703}
\bibinfo{author}{Ye, Y.}, \bibinfo{author}{Tang, T.}, \bibinfo{author}{Zhu, B.}, \bibinfo{author}{Yang, C.}, \bibinfo{author}{Li, B.}, \bibinfo{author}{Hao, S.}, \bibinfo{year}{2022}.
\newblock \bibinfo{title}{A multiscale framework with unsupervised learning for remote sensing image registration}.
\newblock \bibinfo{journal}{IEEE Transactions on Geoscience and Remote Sensing} \bibinfo{volume}{60}, \bibinfo{pages}{1--15}.
\newblock \DOIprefix\doi{10.1109/TGRS.2022.3167644}.
%Type = Article
\bibitem[{Yu et~al.(2024)Yu, Cao, Zhang, Zhang, Hu, Yu, Yu and Shen}]{yu2024internet}
\bibinfo{author}{Yu, J.}, \bibinfo{author}{Cao, S.Y.}, \bibinfo{author}{Zhang, R.}, \bibinfo{author}{Zhang, C.}, \bibinfo{author}{Hu, J.}, \bibinfo{author}{Yu, Z.}, \bibinfo{author}{Yu, B.}, \bibinfo{author}{Shen, H.l.}, \bibinfo{year}{2024}.
\newblock \bibinfo{title}{Internet: Unsupervised cross-modal homography estimation based on interleaved modality transfer and self-supervised homography prediction}.
\newblock \bibinfo{journal}{arXiv preprint arXiv:2409.17993} .
%Type = Inproceedings
\bibitem[{Zhang et~al.(2020)Zhang, Wang, Liu, Jia, Ye, Wang, Zhou and Sun}]{zhang2020content}
\bibinfo{author}{Zhang, J.}, \bibinfo{author}{Wang, C.}, \bibinfo{author}{Liu, S.}, \bibinfo{author}{Jia, L.}, \bibinfo{author}{Ye, N.}, \bibinfo{author}{Wang, J.}, \bibinfo{author}{Zhou, J.}, \bibinfo{author}{Sun, J.}, \bibinfo{year}{2020}.
\newblock \bibinfo{title}{Content-aware unsupervised deep homography estimation}, in: \bibinfo{booktitle}{European Conference on Computer Vision}, \bibinfo{organization}{Springer}. pp. \bibinfo{pages}{653--669}.
%Type = Inproceedings
\bibitem[{Zhang and Ma(2024)}]{pmlr-v235-zhang24ar}
\bibinfo{author}{Zhang, K.}, \bibinfo{author}{Ma, J.}, \bibinfo{year}{2024}.
\newblock \bibinfo{title}{Sparse-to-dense multimodal image registration via multi-task learning}, in: \bibinfo{editor}{Salakhutdinov, R.}, \bibinfo{editor}{Kolter, Z.}, \bibinfo{editor}{Heller, K.}, \bibinfo{editor}{Weller, A.}, \bibinfo{editor}{Oliver, N.}, \bibinfo{editor}{Scarlett, J.}, \bibinfo{editor}{Berkenkamp, F.} (Eds.), \bibinfo{booktitle}{Proceedings of the 41st International Conference on Machine Learning}, \bibinfo{publisher}{PMLR}. pp. \bibinfo{pages}{59490--59504}.
%Type = Inproceedings
\bibitem[{Zhang et~al.(2025)Zhang, Ma, Cao, Luo, Yu, Chen, Li and Shen}]{zhang2025scpnet}
\bibinfo{author}{Zhang, R.}, \bibinfo{author}{Ma, J.}, \bibinfo{author}{Cao, S.Y.}, \bibinfo{author}{Luo, L.}, \bibinfo{author}{Yu, B.}, \bibinfo{author}{Chen, S.J.}, \bibinfo{author}{Li, J.}, \bibinfo{author}{Shen, H.L.}, \bibinfo{year}{2025}.
\newblock \bibinfo{title}{Scpnet: Unsupervised cross-modal homography estimation via intra-modal self-supervised learning}, in: \bibinfo{booktitle}{European Conference on Computer Vision}, \bibinfo{organization}{Springer}. pp. \bibinfo{pages}{460--477}.
%Type = Inproceedings
\bibitem[{Zhao et~al.(2021)Zhao, Huang and Zhang}]{zhao2021deep}
\bibinfo{author}{Zhao, Y.}, \bibinfo{author}{Huang, X.}, \bibinfo{author}{Zhang, Z.}, \bibinfo{year}{2021}.
\newblock \bibinfo{title}{Deep lucas-kanade homography for multimodal image alignment}, in: \bibinfo{booktitle}{Proceedings of the IEEE/CVF conference on computer vision and pattern recognition}, pp. \bibinfo{pages}{15950--15959}.
%Type = Article
\bibitem[{Zhou et~al.(2024)Zhou, Huang, Yan, Hong, Jia, Chanussot and Li}]{Zhou2024AGS}
\bibinfo{author}{Zhou, M.}, \bibinfo{author}{Huang, J.}, \bibinfo{author}{Yan, K.R.}, \bibinfo{author}{Hong, D.}, \bibinfo{author}{Jia, X.}, \bibinfo{author}{Chanussot, J.}, \bibinfo{author}{Li, C.}, \bibinfo{year}{2024}.
\newblock \bibinfo{title}{A general spatial-frequency learning framework for multimodal image fusion.}
\newblock \bibinfo{journal}{IEEE transactions on pattern analysis and machine intelligence} \bibinfo{volume}{PP}.
\newblock \URLprefix \url{https://api.semanticscholar.org/CorpusID:267779599}.
%Type = Inproceedings
\bibitem[{Zhu et~al.(2024)Zhu, Cao, Hu, Zuo, Yu, Ying, Li and Shen}]{zhu2024mcnet}
\bibinfo{author}{Zhu, H.}, \bibinfo{author}{Cao, S.Y.}, \bibinfo{author}{Hu, J.}, \bibinfo{author}{Zuo, S.}, \bibinfo{author}{Yu, B.}, \bibinfo{author}{Ying, J.}, \bibinfo{author}{Li, J.}, \bibinfo{author}{Shen, H.L.}, \bibinfo{year}{2024}.
\newblock \bibinfo{title}{Mcnet: Rethinking the core ingredients for accurate and efficient homography estimation}, in: \bibinfo{booktitle}{Proceedings of the IEEE/CVF Conference on Computer Vision and Pattern Recognition}, pp. \bibinfo{pages}{25932--25941}.
%Type = Inproceedings
\bibitem[{Zhu et~al.(2017)Zhu, Park, Isola and Efros}]{CycleGAN2017}
\bibinfo{author}{Zhu, J.Y.}, \bibinfo{author}{Park, T.}, \bibinfo{author}{Isola, P.}, \bibinfo{author}{Efros, A.A.}, \bibinfo{year}{2017}.
\newblock \bibinfo{title}{Unpaired image-to-image translation using cycle-consistent adversarial networks}, in: \bibinfo{booktitle}{Computer Vision (ICCV), 2017 IEEE International Conference on}.

\end{thebibliography}
\end{sloppypar}
\end{document}